\documentclass[a4]{article}
\usepackage[T1]{fontenc}
\usepackage[utf8]{inputenc}
\usepackage{amsmath, amssymb, amsthm}
\usepackage[english]{babel}
\usepackage{indentfirst}
\usepackage{booktabs}
\usepackage{array}
\usepackage{caption,appendix}
\usepackage{geometry}
\usepackage{float}
\usepackage{cancel}
\usepackage{bbold}
\usepackage{authblk}
\usepackage{cite}
\usepackage{lscape}
\allowdisplaybreaks
\begin{document}
\author{Maurizio Capriolo \thanks{mcapriolo@unisa.it}}
\affil{\emph{Dipartimento di Fisica \lq\lq E. R. Caianiello\rq\rq, Universit\`a degli Studi di Salerno,  via Giovanni Paolo II, 132, Fisciano, SA I-84084, Italy.}}
\affil{\emph{Istituto Nazionale di Fisica Nucleare, Sezione di Napoli, 
Compl. Univ. di Monte S. Angelo, Edificio G, Via Cinthia, I-80126, Napoli, Italy.}}
\title{Gravitational radiation in \\ higher order non-local gravity}
\maketitle

\begin{abstract}
In this paper we examine gravitational radiation in higher order non-local gravity described by the non-local gravitational Lagrangian density \mbox{$\mathcal{L}_{g}=R+\sum_{h=1}^{n}a_{h}R\Box^{-h}R$}.  This non-local theory of gravitation always exhibits the tensor transverse gravitational radiation for $k_{1}^{2}=0$, corresponding to the angular frequency $\omega_{1}$,  composed of two standard $(+)$ and $(\times)$ polarization modes,  massless and of helicity 2.  Furthermore, it shows, under suitable constraint and $n\geq 2$,  an additional massive transverse scalar gravitational radiation with helicity 0. It is composed of $n-1$ modes associated to $n-1$  angular frequencies $\omega_{2},\ldots,\omega_{n}$, each of which of breathing polarization $(b)$ to lowest order in $\gamma$,  a parameter that takes into account the difference in speed between the slightly massive wave and the massless one.  Thanks to NP~formalism, we find that the $E(2)$ class of non-local gravitational waves is $N_{3}$, according Petrov classification, where the presence or absence of all modes are observer independent.  Also, the scalar radiation is forbidden for $n=1$ and $n=2$ cases, when some conditions are satisfied. Finally, in $\Box^{-1}$~gravity where $n=1$,  a possible degenerate case with a continuous infinity of transverse massive scalar breathing modes appears under a particular constraint, which reproduces in two-dimensional spacetime the Polyakov effective action. 
\end{abstract}



\section{Introduction}	

Local theories of gravity are described by Lagrangians that are formed by a finite sum of products between fields and their derivatives  evaluated at the same point $x$ of the spacetime and dynamical variables are governed by differential equations, as in general relativity. Instead, non-local theories of gravity are described by Lagrangians composed by a finite sum of products between fields and their derivatives evaluated at different points $x$ and $x^{\prime}$ of the spacetime and dynamical variables are governed by integro-differential equations. That is, the value of the field at one point depends on its value at another point of the spacetime, weighted by a function called nucleus or kernel. The non-locality in theories of gravity can essentially enter through three different ways. The first manner is by means of a convergent series expansion with real coefficients of an analytic non-polynomial function $\mathcal{F}$ of operator $\Box$,  known as Infinite Derivative of Gravity (IDG)~\cite{BLP,BLMTY,BLY, BGLM,BCHKLM,BLM,BKLM, EF}
\begin{equation}
\mathcal{F}(\Box)=\sum_{n=1}^{\infty}c_{n}\Box^{n}\ .
\end{equation}
In the second way, the non-linearity intervenes in the non-analytic operators as 
\begin{equation}
\Box^{-n}R(x)=\Phi_{0}(x)+\int\,d^{4}x^{\prime}\,\sqrt{-g(x^{\prime})}G(x,x^{\prime})R(x^{\prime})\ ,
\end{equation}
where $G(x,x^{\prime})$ is the retarded Green function of operator $\Box^{n}$, that is 
\begin{equation}
\sqrt{-g(x)}\Box_{x}^{n}G(x,x^{\prime})=\delta(x-x^{\prime})\ ,
\end{equation}
subject to retarded boundary conditions for the principle of causality 
\begin{equation}
G(x,x^{\prime})=0\quad\forall\,t<t^{\prime}\ ,
\end{equation}
and $\Phi_{0}$ is the homogeneous solution of following partial differential equation
\begin{equation}
\Box^{n}\Phi_{0}(x)=0\ .
\end{equation}
Finally, in a different approach where non-locality enters through a non-local constitutive law as in the electrodynamics of media (for details see~\cite{MOP,MNLT2}),  where the constitutive relations between $H_{\mu\nu}\rightarrow(\mathbf{D},\mathbf{H})$ and $F_{\mu\nu}\rightarrow(\mathbf{E}, \mathbf{B})$ involving both the memory of the electromagnetic field, that is, the temporal dispersion and the anisotropy and non-homogeneity of the medium, that is, the spatial dispersion, take the non-local form~\cite{LLECM,JED}
\begin{equation}
H_{\alpha\beta}(x)=\int\,d^{4}x^{\prime}K_{\alpha\beta}^{\phantom{\alpha\beta}\mu\nu}(x^{\prime})F_{\mu\nu}(x-x^{\prime})\ .
\end{equation}

Non-locality could be a manifestation of quantum nature of gravity.  Non-local models of gravity, built by adding to Einstein-Hilbert action, non-local terms like $\Box^{-n}$ or $\mathcal{F}(\Box)$ represent quantum corrections to classical general relativity (loop corrections), can be regarded as a quantum effective field theories.  These models, today, are widely used in cosmology and astrophysics, for example,  to explain the current cosmic expansion or the early times acceleration as well as structure formation,
without introducing dark energy and dark matter. Specifically,  non-locality could play a crucial role to address problems like cosmological constant, Big Bang and black hole singularities, and, in general, coincidence and fine-tuning problems, which affect the $\Lambda$CDM model~\cite{CBNLC, MR, BCSU,CALC}. Also, these effective theories are often ghost-free and stable,  consistent with Solar system constraints, super-renormalizability at quantum level and then less affected by  infrared and ultraviolet divergences~\cite{Modesto2}.

In this paper we analyze gravitational waves~(GWs) in non-local theory of gravity described by Lagrangian density of higher order, linear in $R$ and $\Box^{-h}$, i.e.  $\mathcal{L}_{g}=R+\sum_{h=1}^{n}a_{h}R\Box^{-h}R$.  In particular, we want to investigate how non-local terms can contribute to adding a further modes in gravitational radiation, different from the transverse tensor ones, predicted in general relativity.  

This paper is organized as follows. In section~\ref{sec2},  we localize the non-local action through the introduction of suitable scalar fields, i.e.  via Lagrangian multipliers approach, deriving the related field equations.  Then, in section~\ref{sec3} weakly perturbing the metric tensor and scalar fields we obtain the linearized equations and  in suitable gauge we find their wavelike solutions.  Sections~\ref{sec4} and \ref{sec5} are devoted to study of polarizations and helicities via geodesic deviation and the Newman–Penrose (NP) formalism.  Below,  the section~\ref{table} provides some tables while results and final remarks are summarized in section~\ref{finalremarks}.
\section{Localization of higher order non-local action via Lagrangian multipliers approach}\label{sec2}
We want to study the gravity governed by the following higher order non-local action
\begin{equation}\label{1}
S[g]=\frac{1}{2\kappa^2}\int d^4x \sqrt{-g} \biggl(R+\sum_{h=1}^{n}a_{h}R\Box^{-h}R\biggr)+\int d^4x \sqrt{-g}\mathcal{L}_{m}[g]\ ,
\end{equation}
where $\kappa^{2}=8\pi G/c^{4}$ and the D'Alembert operator $\Box$, is defined as $\Box=g^{\mu\nu}\nabla_{\mu}\nabla_{\nu}$.  Its field equations are non-linear integro-differential equations hard to solve and therefore a different procedure is used to analyze this model of non-local gravity,  starting from an action equivalent to that of Eq.~\eqref{1}.  Thus,  we introduce $n$ auxiliary fields $\phi_{1}(x),\dots,\phi_{n}(x)$ as 
\begin{align}\label{2}
\phi_{1}(x)&=\Box^{-1}R(x)\ ,\\
\phi_{2}(x)&=\Box^{-2}R(x)\ ,\\
&\vdots\notag\\
\phi_{n}(x)&=\Box^{-n}R(x)\ ,
\end{align}
involving
\begin{align}\label{3}
R(x)&=\Box\phi_{1}(x)\ ,\\
\phi_{1}(x)&=\Box\phi_{2}\left(x\right)\ ,\\
\vdots\\
\phi_{n-1}(x)&=\Box\phi_{n}\left(x\right)\ ,
\end{align}
and $n$ Lagrange multipliers,  new scalar fields $\lambda_{1}(x),\dots,\lambda_{n}(x)$,  so that the free gravitational action Eq.~\eqref{1} is equivalent to~\cite{NO, DWCI, DWCII, DW3}
\begin{multline}\label{4}
S_{g}[g,\phi_{1},\dots\phi_{n},\lambda_{1},\dots,\lambda_{n}]=\frac{1}{2\kappa^2}\int d^4x \sqrt{-g}\Bigg[R\biggl(1+\sum_{h=1}^{n}a_{h}\phi_{h}\biggr)\\
+\lambda_{1}\left(R-\Box\phi_{1}\right)+\sum_{h=1}^{n-1}\lambda_{h+1}\left(\phi_{h}-\Box\phi_{h+1}\right)\Bigg]\ .
\end{multline}
Using the integration by parts,  the Gauss theorem and imposing that fields and their derivatives vanish on the boundary of integration domain, we get
\begin{multline}\label{5}
S_{g}[g,\phi_{1},\dots\phi_{n},\lambda_{1},\dots,\lambda_{n}]=\frac{1}{2\kappa^2}\int d^4x \sqrt{-g} \Bigg[\biggl(1+\sum_{h=1}^{n}a_{h}\phi_{h}+\lambda_{1}\biggr)R\\
+\sum_{h=1}^{n-1}\lambda_{h+1}\phi_{h}+\sum_{h=1}^{n}\nabla^{\nu}\lambda_{h}\nabla_{\nu}\phi_{h}\Bigg]\ .
\end{multline}
The variational derivative of the gravitational action Eq.~\eqref{4} with respect to $\phi_{h}$ and $\phi_{n}$ with $1\leq h\leq n-1$, yields
\begin{equation}\label{6}
\frac{2\kappa^{2}}{\sqrt{-g}}\frac{\delta S_{g}}{\delta\phi_{h}}=a_{h}R+\lambda_{h+1}-\Box\lambda_{h},
\end{equation}
\begin{equation}\label{6_1}
\frac{2\kappa^{2}}{\sqrt{-g}}\frac{\delta S_{g}}{\delta\phi_{n}}=a_{n}R-\Box\lambda_{n},
\end{equation}
while the functional derivative  with respect to $\lambda_{1}$ and $\lambda_{l}$,  with $2\leq l\leq n$, takes the following form 
\begin{equation}\label{7}
\frac{2\kappa^{2}}{\sqrt{-g}}\frac{\delta S_{g}}{\delta\lambda_{1}}=R-\Box\phi_{1}\ ,
\end{equation}
\begin{equation}\label{7_1}
\frac{2\kappa^{2}}{\sqrt{-g}}\frac{\delta S_{g}}{\delta\lambda_{l}}=\phi_{l-1}-\Box\phi_{l}\ .
\end{equation}
Finally,  varying with respect to $g^{\mu\nu}$ the gravitational and the material part of the action, Eq.~\eqref{4}  and Eq.~\eqref{1}, we have, respectively, the following functional derivatives
\begin{align}\label{8}
\frac{2\kappa^{2}}{\sqrt{-g}}\frac{\delta S_{g}}{\delta g^{\mu\nu}}&=\left(G_{\mu\nu}+g_{\mu\nu}\Box-\nabla_{\mu}\nabla_{\nu}\right)\biggl(1+\sum_{h=1}^{n}a_{h}\phi_{h}+\lambda_{1}\biggr)-\frac{1}{2}g_{\mu\nu}\sum_{h=1}^{n-1}\phi_{h}\lambda_{h+1}\nonumber\\
&+\sum_{h=1}^{n}\nabla_{(\mu}\phi_{h}\nabla_{\nu)}\lambda_{h}-\frac{1}{2}g_{\mu\nu}\sum_{h=1}^{n}\nabla^{\sigma}\phi_{h}\nabla_{\sigma}\lambda_{h}\ ,
\end{align}
and 
\begin{equation}\label{9}
\frac{\delta S_{m}}{\delta g^{\mu\nu}}=-\frac{\sqrt{-g}}{2}T_{\mu\nu}\ .
\end{equation}
Thanks to Eqs.~\eqref{6}, \eqref{7}, \eqref{8} and \eqref{9} we derive a second order non-linear field equation in matter,  that is
\begin{align}\label{eqfieldginmatter}
\left(G_{\mu\nu}+g_{\mu\nu}\Box-\nabla_{\mu}\nabla_{\nu}\right)&\biggl(1+\sum_{h=1}^{n}a_{h}\phi_{h}+\lambda_{1}\biggr)-\frac{1}{2}g_{\mu\nu}\sum_{h=1}^{n-1}\phi_{h}\lambda_{h+1}\nonumber\\
&+\sum_{h=1}^{n}\nabla_{(\mu}\phi_{h}\nabla_{\nu)}\lambda_{h}-\frac{1}{2}g_{\mu\nu}\sum_{h=1}^{n}\nabla^{\sigma}\phi_{h}\nabla_{\sigma}\lambda_{h}=\kappa^{2}T_{\mu\nu}
\ ,
\end{align}
plus $2n$ non-linear differential equations each of them of $2$-th degree 
\begin{align}\label{eqfieldphi}
\Box\phi_{1}&=R\ ,\notag\\
&\vdots\notag\\
\Box\phi_{l}&=\phi_{l-1}\ ,\notag\\
&\vdots\\
\Box\lambda_{h}&=a_{h}R+\lambda_{h+1}\ ,\notag\\
&\vdots\notag\\
\Box\lambda_{n}&=a_{n}R\ ,\notag
\end{align}
with $2\leq l\leq n$ and $ 1\leq h\leq n-1$, where $G_{\mu\nu}$ is the Einstein tensor
\begin{equation}\label{10.1}
G_{\mu\nu}=R_{\mu\nu}-\frac{1}{2}g_{\mu\nu}R\ .
\end{equation}
Thus, the order of theory is $4n+2$ for $\Box^{-n}$~gravity.  Thereafter, we perform the trace of Eq.~\eqref{eqfieldginmatter} that yields  the trace equation
\begin{equation}\label{eqtrace0}
%\boxed{
\left(R-3\Box\right)\biggl(1+\sum_{h=1}^{n}a_{h}\phi_{h}+\lambda_{1}\biggr)+2\sum_{h=1}^{n-1}\phi_{h}\lambda_{h+1}+\sum_{h=1}^{n}\nabla^{\alpha}\lambda_{h}\nabla_{\alpha}\phi_{h}=-\kappa^{2}T
%}
\ .
\end{equation}
\section{Linearization and gravitational waves}\label{sec3}
We investigate the gravitational waves at very large distances from their sources, where the weak field approximation is applicable.  So, in order to analyze radiation in non-local gravitational theories,  we first perturb the metric tensor $g_{\mu\nu}$ around the flat metric $\eta_{\mu\nu}$ and the $2n$ scalar fields $\phi_{1},\ldots,\phi_{n}$ and $\lambda_{1},\ldots,\lambda_{n}$ around their constant values assumed in the Minkowskian spacetime, $\phi^{(0)}$ and $\lambda^{(0)}$~\cite{WAG},  respectively as 
\begin{equation}\label{11}
\begin{aligned}
&&g_{\mu\nu}&\sim\eta_{\mu\nu}+h_{\mu\nu}\ ,\\ 
&&\phi_{1}&\sim\phi_{1}^{(0)}+\delta\phi_{1}\ ,\\
&&&\vdots\\
&&\phi_{n}&\sim\phi_{n}^{(0)}+\delta\phi_{n}\ ,\\ 
&&\lambda_{1}&\sim\lambda_{1}^{(0)}+\delta\lambda_{1}\ ,\\
&&&\vdots\\ 
&&\lambda_{n}&\sim\lambda_{n}^{(0)}+\delta\lambda_{n}\ .
\end{aligned}
\end{equation}
Then, to first order in $h_{\mu\nu}$, the Riemann tensor $R_{\mu\nu}$, the Ricci scalar $R$ and the Einstein tensor $G_{\mu\nu}$ become 
\begin{equation}\label{12}
R_{\mu\nu}^{(1)}=\frac{1}{2}\left(\partial_{\sigma}\partial_{\mu}h^{\sigma}_{\nu}+\partial_{\sigma}\partial_{\nu}h^{\sigma}_{\mu}-\partial_{\mu}\partial_{\nu}h-\Box h_{\mu\nu}\right)\ ,
\end{equation}
\begin{equation}\label{13}
R^{(1)}=\partial_{\mu}\partial_{\nu}h^{\mu\nu}-\Box h\ ,
\end{equation}
\begin{equation}\label{13_1}
G_{\mu\nu}^{(1)}=\frac{1}{2}\left(\partial_{\sigma}\partial_{\mu}h^{\sigma}_{\nu}+\partial_{\sigma}\partial_{\nu}h^{\sigma}_{\mu}-\partial_{\mu}\partial_{\nu}h-\Box h_{\mu\nu}-\eta_{\mu\nu}\partial_{\alpha}\partial_{\beta}h^{\alpha\beta}+\eta_{\mu\nu}\Box h\right)\ ,
\end{equation}
where $h$ is the trace of perturbation $h_{\mu\nu}$ and $\Box=\eta^{\mu\nu}\partial_{\mu}\partial_{\nu}$ is the D'Alembert operator in the Minkowskian space.
Later,  according to Eqs.~\eqref{eqfieldginmatter},  \eqref{eqfieldphi},  \eqref{eqtrace0},  \eqref{11}, \eqref{12}, \eqref{13} and \eqref{13_1}, the linearized field equations up to first order are 
\begin{multline}\label{Lineareqfieldginmatter}
\biggl(1+\sum_{h=1}^{n}a_{h}\phi_{h}+\lambda_{1}\biggr)^{(0)}G_{\mu\nu}^{(1)}+\Bigl(g_{\mu\nu}\Box-\nabla_{\mu}\nabla_{\nu}\Bigr)^{(0)}\biggl(1+\sum_{h=1}^{n}a_{h}\phi_{h}+\lambda_{1}\biggr)^{(1)}\\
-\frac{1}{2}g_{\mu\nu}^{(1)}\sum_{h=1}^{n-1}\phi_{h}^{(0)}\lambda_{h+1}^{(0)}-\frac{1}{2}g_{\mu\nu}^{(0)}\sum_{h=1}^{n-1}\phi_{h}^{(1)}\lambda_{h+1}^{(0)}-\frac{1}{2}g_{\mu\nu}^{(0)}\sum_{h=1}^{n-1}\phi_{h}^{(0)}\lambda_{h+1}^{(1)}\\
-\frac{1}{2}g_{\mu\nu}^{(0)}\sum_{h=1}^{n-1}\phi_{h}^{(0)}\lambda_{h+1}^{(0)}=\kappa^{2}T_{\mu\nu}^{(0)}
\ ,
\end{multline}
\begin{equation}\label{14}
\begin{gathered}
\Box\delta\phi_{1}=R^{(1)}\ ,\\
\vdots\\
\Box\delta\phi_{l}=\phi_{l-1}^{(0)}+\delta\phi_{l-1}\ ,\\
\vdots\\
\Box\delta\lambda_{h}=a_{h}R^{(1)}+\lambda_{h+1}^{(0)}+\delta\lambda_{h+1}\ ,\\
\vdots\\
\Box\delta\lambda_{n}=a_{n}R^{(1)}\ ,
\end{gathered}
\end{equation}
\begin{multline}\label{eqtrace}
\biggl(1+\sum_{h=1}^{n}a_{h}\phi_{h}+\lambda_{1}\biggr)^{(0)}R^{(1)}-3\Box^{(0)}\biggl(1+\sum_{h=1}^{n}a_{h}\phi_{h}+\lambda_{1}\biggr)^{(1)}\\
+2\sum_{h=1}^{n-1}\phi_{h}^{(0)}\lambda_{h+1}^{(1)}+2\sum_{h=1}^{n-1}\phi_{h}^{(1)}\lambda_{h+1}^{(0)}+2\sum_{h=1}^{n-1}\phi_{h}^{(0)}\lambda_{h+1}^{(0)}=-\kappa^{2}T^{(0)}
\ ,
\end{multline}
where $2\leq l\leq n$ and $1\leq h\leq n-1$. Equaling terms to zero-th order on the left and right of Eqs.~\eqref{14},  we obtain
\begin{equation}
\phi_{l-1}^{(0)}=0\quad \text{and}\quad\lambda_{h+1}^{(0)}=0\ .
\end{equation}
Thus, field equations and the trace become 
\begin{equation}\label{Lineareqfieldginmatter1}
\Bigl(1+a_{n}\phi_{n}^{(0)}+\lambda_{1}^{(0)}\Bigr)G_{\mu\nu}^{(1)}+\Bigl(\eta_{\mu\nu}\Box-\partial_{\mu}\partial_{\nu}\Bigr)\biggl(\sum_{h=1}^{n}a_{h}\delta\phi_{h}+\delta\lambda_{1}\biggr)=\kappa^{2}T_{\mu\nu}^{(0)}
\ ,
\end{equation}
\begin{equation}\label{15}
\begin{aligned}
\Box\delta\phi_{1}&=R^{(1)}\ ,\\
\vdots\\
\Box\delta\phi_{l}&=\delta\phi_{l-1}\ ,\\
&\vdots&\\
\Box\delta\lambda_{h}&=a_{h}R^{(1)}+\delta\lambda_{h+1}\ ,\\
\vdots\\
\Box\delta\lambda_{n}&=a_{n}R^{(1)}\ ,
\end{aligned}
\end{equation}
\begin{equation}\label{eqtrace1}
%\boxed{
\Bigl(1+a_{n}\phi_{n}^{(0)}+\lambda_{1}^{(0)}\Bigr)R^{(1)}-3\Box\biggl(\,\sum_{h=1}^{n}a_{h}\delta\phi_{h}+\delta\lambda_{1}\biggr)=-\kappa^{2}T^{(0)}
%}
\ .
\end{equation}
Now, we define a new gauge in a coordinate frame $\left\{x^{\mu}\right\}$ with $1+a_{n}\phi_{n}^{(0)}+\lambda_{1}^{(0)}\neq 0$ as
\begin{equation}\label{100}
\bar{h}_{\mu\nu}=h_{\mu\nu}-\frac{1}{2}\eta_{\mu\nu}h-\frac{\eta_{\mu\nu}}{1+a_{n}\phi_{n}^{(0)}+\lambda_{1}^{(0)}}\biggl(\,\sum_{h=1}^{n}a_{h}\delta\phi_{h}+\delta\lambda_{1}\biggr)\ ,
\end{equation}
such that
\begin{equation}\label{101}
\partial_{\mu}\bar{h}^{\mu\nu}=0\ .
\end{equation}
The trace of Eq.~\eqref{100} is 
\begin{equation}\label{102}
\bar{h}=-h-\frac{4}{1+a_{n}\phi_{n}^{(0)}+\lambda_{1}^{(0)}}\biggl(\,\sum_{h=1}^{n}a_{h}\delta\phi_{h}+\delta\lambda_{1}\biggr)\ .
\end{equation}
In terms of barred quantities $\bar{h}_{\mu\nu}$ and $\bar{h}$,  we obtain from Eqs.~\eqref{13},\eqref{13_1}, \eqref{100} and \eqref{102}
\begin{equation}\label{103}
R^{(1)}=\frac{1}{2}\Box\bar{h}+\frac{3}{1+a_{n}\phi_{n}^{(0)}+\lambda_{1}^{(0)}}\Box\biggl(\,\sum_{h=1}^{n}a_{h}\delta\phi_{h}+\delta\lambda_{1}\biggr)\ ,
\end{equation}
\begin{equation}\label{104}
G_{\mu\nu}^{(1)}=-\frac{1}{2}\Box\bar{h}_{\mu\nu}-\frac{1}{1+a_{n}\phi_{n}^{(0)}+\lambda_{1}^{(0)}}\left(\eta_{\mu\nu}\Box-\partial_{\mu}\partial_{\nu}\right)\biggl(\,\sum_{h=1}^{n}a_{h}\delta\phi_{h}+\delta\lambda_{1}\biggr)\ .
\end{equation}
If we set $B=1+a_{n}\phi_{n}^{(0)}+\lambda_{1}^{(0)}\neq 0$,  field equations can be rewritten as 
\begin{equation}\label{105}
\Box\bar{h}_{\mu\nu}=-\frac{2\kappa^{2}}{B}T_{\mu\nu}^{(0)}\ ,
\end{equation}
\begin{equation}\label{106}
\begin{aligned}
%\begin{split}
\Box\delta\phi_{1}&=\frac{1}{2}\Box\bar{h}+\frac{3}{B}\biggl(\,\sum_{h=1}^{n}a_{h}\Box\delta\phi_{h}+\Box\delta\lambda_{1}\biggr)\ ,\\
&\vdots\\
\Box\delta\phi_{l}&=\delta\phi_{l-1}\ ,\\
&\vdots\\
\Box\delta\lambda_{h}&=\frac{a_{h}}{2}\Box\bar{h}+\frac{3a_{h}}{B}\Box\biggl(\,\sum_{l=1}^{n}a_{l}\delta\phi_{l}+\delta\lambda_{1}\biggr)+\delta\lambda_{h+1}\ ,\\
&\vdots\\
\Box\delta\lambda_{n}&=\frac{a_{n}}{2}\Box\bar{h}+\frac{3a_{n}}{B}\Box\biggl(\,\sum_{h=1}^{n}a_{h}\delta\phi_{h}+\delta\lambda_{1}\biggr)\ ,
%\end{split}
\end{aligned}
\end{equation}
that is, ten linear differential equations in $\bar{h}_{\mu\nu}$ and $2n$ linear differential equations in $\delta\phi_{1},\ldots,\delta\phi_{n}$ and $\delta\lambda_{1},\ldots,\delta\lambda_{n}$.
While, finally, the linearized equation of trace becomes 
\begin{equation}\label{107}
\Box \bar{h}=-\frac{2\kappa^{2}}{B}T^{(0)}\ .
\end{equation}
\subsection{The propagation of gravitational waves}
The study of only propagation of waves starts from the linearized field equations in vacuum, that is, in absence of massive objects, because we are not interested in their generation. Hence,  Eqs.~\eqref{105} and \eqref{106} take the form
\begin{equation}\label{199}
\Box\bar{h}_{\mu\nu}=0\ ,
\end{equation}
\begin{gather}
%\begin{split}
\left(3a_{1}-B\right)\Box\delta\phi_{1}+3\sum_{j=2}^{n}a_{j}\Box\delta\phi_{j}+3\Box\delta\lambda_{1}=0\ ,\notag\\
\vdots\notag\\
\Box\delta\phi_{l}-\delta\phi_{l-1}=0\ ,\notag\\
\vdots\label{linearizedsistem}\\
a_{h}\Box\delta\phi_{1}-\Box\delta\lambda_{h}+\delta\lambda_{h+1}=0\ ,\notag\\
\vdots\notag\\
a_{n}\Box\delta\phi_{1}-\Box\delta\lambda_{n}=0\ ,\notag
%\end{split}\ ,
\end{gather}
where $l$ range over $2,\dots, n$ and $h$ over $1,\dots,n-1$.
Adopting the following Fourier decomposition 
\begin{equation}\label{200}
h_{\mu\nu}\left(x\right)=\frac{1}{\left(2\pi\right)^{3/2}}\int d^{3}\mathbf{k}\,\tilde{h}_{\mu\nu}\left(\mathbf{k}\right)e^{i k\cdot x}\ ,
\end{equation}
and the same decomposition for the other $2n$ scalar fields, we get
\begin{equation}\label{201}
k^{2}\tilde{\bar{h}}_{\mu\nu}=0\ ,
\end{equation}
\begin{gather}
k^{2}\left[\left(3a_{1}-B\right)\widetilde{\delta\phi}_{1}+3\sum_{j=2}^{n}a_{j}\widetilde{\delta\phi}_{j}+3\widetilde{\delta\lambda}_{1}\right]=0\ ,\notag\\
\vdots\notag\\
\widetilde{\delta\phi}_{l-1}+k^{2}\widetilde{\delta\phi}_{l}=0\ ,\notag\\
\vdots\label{linearizedsistemFspace}\\
a_{h}k^{2}\widetilde{\delta\phi}_{1}-k^{2}\widetilde{\delta\lambda}_{h}-\widetilde{\delta\lambda}_{h+1}=0\ ,\notag\\
\vdots\notag\\
k^{2}\left(a_{n}\widetilde{\delta\phi}_{1}-\widetilde{\delta\lambda}_{n}\right)=0\ .\notag
\end{gather}
If we consider a null plane wave $k^{2}=0$, we get
\begin{equation}\label{203}
\widetilde{\delta\phi}_{l-1}=\widetilde{\delta\lambda}_{h+1}=0,\quad\text{with}\quad l=2,\dots, n \quad \text{and}\quad h=1,\dots,n-1\ ,
\end{equation}
and 
\begin{equation}\label{203.1}
a_{n}\widetilde{\delta\phi}_{n}+\widetilde{\delta\lambda}_{1}=0\ .
\end{equation}
Therefore, the linear metric perturbation $h_{\mu\nu}$ Eq.~\eqref{100} is reduced to the well-know relation 
\begin{equation}\label{203.2}
\bar{h}_{\mu\nu}=h_{\mu\nu}-\frac{1}{2}\eta_{\mu\nu}h\ .
\end{equation}
Then, in an suitable gauge that leaves invariant $\partial_{\mu}\bar{h}^{\mu\nu}=0$,  that is with $\Box\xi^{\nu}=0$ under infinitesimal transformation $x^{\prime\mu}=x^{\mu}+\xi^{\mu}$, the traceless tensor gravitational wave with $k_{1}^{2}=0$ is 
\begin{equation}\label{204}
h_{\mu\nu}^{(k_{1})}\left(x\right)=\frac{1}{\left(2\pi\right)^{3/2}}\int d^{3}\mathbf{k}\,\widetilde{C}_{\mu\nu}\left(\mathbf{k}\right)e^{i k_{1}\cdot x}+c.c.\ ,
\end{equation}
where we have used the four-vector $k_{1}^{\mu}=(\omega_{1},\mathbf{k})$ with dispersion relation 
\begin{equation}\label{204.1}
\omega_{1}=\vert \mathbf{k}\vert\ ,
\end{equation}
and $\widetilde{C}_{\mu\nu}$ is the transverse and tracefree polarization tensor in the momentum space, i.e. $\widetilde{C}^{\sigma}_{\phantom{\sigma}\sigma}=0$ and $ k^{\mu}\widetilde{C}_{\mu\nu}=0$. 
If we explore a non-null plane wave $k^{2}\neq 0$, in addition to the trivial solution, we have from Eq.~\eqref{201}
\begin{equation}\label{205}
\tilde{\bar{h}}_{\mu\nu}\left(\mathbf{k}\right)=0\Rightarrow\tilde{\bar{h}}\left(\mathbf{k}\right)=0\ ,
\end{equation}
and coefficients square matrix $A_{2n}$ of the linear homogeneous system in $\widetilde{\delta\phi}_{1}$, $\widetilde{\delta\phi}_{l}$, $\widetilde{\delta\lambda}_{h}$ and $\widetilde{\delta\lambda}_{n}$ Eq.~\eqref{linearizedsistemFspace} is given by 
\begin{equation}\label{206}
A_{2n}=
\left(
\begin{array}{cccccccccccc}
3a_{1}-B		&	3a_{2}	&	3a_{3}	 	&\cdots& 3a_{n-1}		&	3a_{n}			&	3			&	0				&	0 				&\cdots&	0 			& 0\\
1					&	k^{2}	&		0			&\cdots& 	 0		&		0		&	0			&	0				&	0				&\cdots&	0 			&	0\\
0					&		1		&	k^{2}		&\cdots& 	 0		&		0		&	0			&	0				&	0				&\cdots&	0 			&	0\\
\vdots			&	\vdots	&	\vdots		&	\vdots&		\vdots	&	\vdots	&	\vdots	&	\vdots		&	\vdots		&	\vdots&		\vdots	&	\vdots\\
0					&	0			&	0				&\cdots&	1			&	k^{2}	&	0			&	0				&	0				&\cdots	&	0			&	0\\
a_{1}k^{2}		&	0			&	0				&\cdots& 	0		&	0			&	-k^{2}	&	-1				&	0				&\cdots	&	0 			&	0\\
a_{2}k^{2}	&	0			&	0				&\cdots& 	0		&	0			&	0			&	-k^{2}		&	-1				&\cdots	&	0 			&	0\\
\vdots			&	\vdots	&	\vdots		&	\vdots	&		\vdots &	\vdots	&	\vdots	&	\vdots		&	\vdots		& \vdots&	\vdots			&	\vdots\\
a_{n-1}k^{2}	&	0 			&	0				&\cdots& 	0		&	0			&	0			&	0				&	0				&\cdots& -k^{2} 	& -1\\
a_{n}				&	0			&	0				&\cdots& 	0		&	0			&	0			&	0				&	0				&\cdots&0			& -1
\end{array}
\right)\ .
\end{equation}
Its determinant is 
\begin{multline}\label{algebraiceq}
\lvert A_{2n}\rvert=\left(-1\right)^{n}k^{2\left(n-1\right)}\biggl[\left(6a_{1}-B\right)k^{2\left(n-1\right)}\\+\theta(n-2)\sum_{j=2}^{n}6\left(-1\right)^{j+1}a_{j}k^{2\left(n-j\right)}\biggr]\ ,
\end{multline}
where $\theta(n-2)$ is Heaviside function that vanishes for $n< 2$ and equals to $1$ otherwise. The square bracket in Eq.~\eqref{algebraiceq} is a $(n-1)$-th degree polynomial in $k^2\neq 0$ , that vanishes if
\begin{equation}\label{207}
\vert A_{2n}\vert=0\quad\Leftrightarrow\quad\left(6a_{1}-B\right)\prod_{j=2}^{n}\left(k^{2}-k_{j}^{2}\right)=0\ ,
\end{equation}
where $a_{j}$ are such that the equation $\lvert A_{2n}\rvert=0$ admits $n-1$ positive real solutions $k^{2}_{j}\in \mathbb{R}_{+}$. If the next constraint is fulfilled
\begin{equation}\label{constraintBeq0}
B-6a_{1}=0\ ,
\end{equation}
the determinant Eq.~\eqref{algebraiceq} becomes a $(n-2)$-th degree polynomial in $k^{2}\neq 0$
\begin{equation}\label{207.01}
\vert A_{2n}\vert=6\left(-1\right)^{n}k^{2\left(n-1\right)}\left[\theta(n-2)\sum_{j=2}^{n}\left(-1\right)^{j+1}a_{j}k^{2\left(n-j\right)}\right]\ .
\end{equation}
Non-zero roots of the polynomial~\eqref{207.01} in $k^{2}$ under constraint Eq.~\eqref{constraintBeq0}, varying $n$ and under suitable conditions for coefficients $a_{j}$, are
\begin{align}\label{207.1} 
&n=1&&\vert A_{2}\vert=0& &k^{2}\in \mathbb{R}_{+}\;\text{solutions}\nonumber\\
&&&&&\text{there are $\infty^{1}$ solutions}\ ,\\
&n=2&&\vert A_{4}\vert=-6a_{2}k^{2}=0& &\text{there is no solutions} , \\
&n=3&&\vert A_{6}\vert=-6k^{4}\left[-a_{2}k^{2}+a_{3}\right]=0&&\text{one solution $k_{2}^{2}$}\ ,\\
&n=4&&\vert A_{8}\vert=6 k^{6}\left[-a_{2}k^{4}+a_{3}k^{2}-a_{4}\right]=0& &\text{two solutions $k_{2}^{2}$ and $k_{3}^{2}$}\ ,\\
&n&&\vert A_{2n}\vert=0&&\text{there are $n-2$ solutions}\nonumber\\
&&&&&k_{2}^{2},\ldots,k_{n-1}^{2}\ .
\end{align}
%and $k^{2}\in R^{+}$ we have $\infty^{1}$ solutions, that is there are infinite scalar modes but only one polarization as we see after. 
While for the following other constraint 
\begin{equation}\label{constraintBneq0}
B-6a_{1}\neq 0\  ,
\end{equation}
and $k^{2}=k_{j}^{2}=M_{j}^{2}\neq 0$ with $j=2,\dots,n$, we have $n-1$ solutions for  $n\geq 2$ and no solution like $k^{2}\neq 0$ for $n=1$. From the relation between the traces $\bar{h}$ and $h$ Eq.~\eqref{102} and the first equation of the linear system Eq.~\eqref{linearizedsistemFspace} in $\mathbf{k}$-space we obtain in $k^{2}\neq 0$ case for which both $\tilde{\bar{h}}_{\mu\nu}\left(\mathbf{k}\right)$ and $\tilde{\bar{h}}\left(\mathbf{k}\right)$ vanish 
\begin{equation}\label{208}
\tilde{h}\left(\mathbf{k}\right)=-\frac{4}{3}\widetilde{\delta\phi}_{1}\left(\mathbf{k}\right)\ ,
\end{equation}
and consequently
\begin{equation}\label{209}
\tilde{h}_{\mu\nu}\left(\mathbf{k}\right)=-\frac{\eta_{\mu\nu}}{3}\widetilde{\delta\phi}_{1}\left(\mathbf{k}\right)\ .
\end{equation}
Then, the $n-1$ non-null wavelike solutions, linked to ${j}$-th mode, are of the type
\begin{equation}\label{210}
h_{\mu\nu}^{(k_{j})}\left(x\right)=\frac{1}{\left(2\pi\right)^{3/2}}\int d^{3}\mathbf{k}\,\frac{\eta_{\mu\nu}}{4}\tilde{A}_{j}\left(\mathbf{k}\right)e^{i k_{j}\cdot x}+c.c.\ ,
\end{equation}
with the four-vector $k_{j}^{\mu}=(\omega_{j},\mathbf{k})$ where the $j$-th amplitude in $\mathbf{k}$ space is 
\begin{equation}\label{210.1}
\tilde{A}_{j}\left(\mathbf{k}\right)=\tilde{h}^{(k_{j})}\left(\mathbf{k}\right)\ ,
\end{equation}
for the $j$-th dispersion relation 
\begin{equation}\label{210.2}
\omega_{j}\left(\mathbf{k}\right)=\sqrt{\vert\mathbf{k}\vert^{2}+M_{j}^{2}}\ .
\end{equation}
To show that $k_{j}$-waves are massive, namely the linear perturbation of scalar fields satisfy an equation like Klein-Gordon, according to the linearized system Eq.~\eqref{linearizedsistem}, we can rewrite it as equivalent system
\begin{gather}
%\begin{split}
\left(3a_{1}-B\right)\Box\delta\phi_{1}+3\sum_{j=2}^{n}a_{j}\Box\delta\phi_{j}+3\Box\delta\lambda_{1}=0\ ,\notag\\
\vdots\notag\\
\Box^{l}\delta\phi_{l}-\Box\delta\phi_{1}=0\ ,\notag\\
\vdots\label{211}\\
a_{h}\Box\delta\phi_{1}-\Box\delta\lambda_{h}+\delta\lambda_{h+1}=0\ ,\notag\\
\vdots\notag\\
\sum_{j=1}^{n}a_{j}\Box^{n-j+1}\delta\phi_{1}-\Box^{n}\delta\lambda_{1}=0\ ,\notag
%\end{split}\ ,
\end{gather}
with $l=2,\dots, n-1$ and $h=1,\dots,n-1$.  Then, by acting of the differential operator $\Box^{n-1}$ on the first equation of the system Eq.~\eqref{211} we get 
\begin{equation}\label{216}
\left[\left(6a_{1}-B\right)\Box^{n}+3\sum_{j=2}^{n}a_{j}\left(\Box^{j-1}+\Box^{n-j+1}\right)\right]\delta\phi_{1}=0\ ,
\end{equation}
that can be decomposed as
\begin{equation}\label{217}
\left[\left(6a_{1}-B\right)\prod_{j=1}^{n}\left(\Box+k_{j}^{2}\right)\right]\delta\phi_{1}=0\ ,
\end{equation}
under the condition that $k_{j}^{2}\neq 0$ when $j$ run over $2,\dots,n$, are roots of $(n-1)$-th degree polynomial Eq.~\eqref{algebraiceq} in $k^{2}$ and $k_{1}^{2}=0$. If we define the new scalar fields $\delta\Phi_{h}$, with $h\in\{2,\dots,n\}$, as~\cite{SCHMIDTVD} 
\begin{equation}\label{218}
\delta\Phi_{h}=\prod_{j=1,j\neq h}^{n}\left(\Box+M_{j}^{2}\right)\delta\phi_{1}\ ,
\end{equation}
we obtain the Klein-Gordon equation for each scalar fields  $\delta\Phi_{h}$  with mass $M_{h}\neq 0$ and $M_{1}=0$, that is
\begin{equation}\label{219}
\left(\Box+M_{h}^{2}\right)\delta\Phi_{h}=0\ ,
\end{equation}
and accordingly
\begin{equation}\label{230}
\delta\Phi_{h}=\frac{\prod_{j=1,j\neq h}^{n}\left(M_{j}^{2}-M_{h}^{2}\right)}{(2\pi)^{3/2}}\int d^{3}\mathbf{k}\widetilde{\delta\phi}_{1}(\mathbf{k})e^{i k_{h}\cdot x}\ .
\end{equation}
In summary, the gravitational wave of higher order non-local gravity,  from superposition principle can be expressed as
\begin{multline}\label{GWsNonLocalVacuum}
h_{\mu\nu}\left(x\right)=\frac{1}{\left(2\pi\right)^{3/2}}\int d^{3}\mathbf{k}\,\widetilde{C}_{\mu\nu}\left(\mathbf{k}\right)e^{i k_{1}\cdot x}\\+\frac{1}{\left(2\pi\right)^{3/2}}\sum_{j=2}^{n}\int d^{3}\mathbf{k}\,\frac{\eta_{\mu\nu}}{4}\tilde{A}_{j}\left(\mathbf{k}\right)e^{i k_{j}\cdot x}+c.c. \ ,
\end{multline}
that is, one massless wave $k_{1}^{2}=0$ and $n-1$ massive waves $k_{j}^{2}=M_{j}^{2}\neq 0$ under the constraint $B\neq 6a_{1}$ occur. It should be noted that when $B=6a_{1}$, massive waves are $n-2$, because one non-null wave is inhibit and the $j$~index in Eq.~\eqref{GWsNonLocalVacuum} run from $2$ to $n-1$. For the study of GWs in higher order local gravity governed by Lagrangian $R+\sum_{h=0}^{n}a_{h}R\Box^{h}R$, see~\cite{CCC}, and for a confront between general relativity and experiments, see~\cite{WGRE,GW17_1,GW17_2,GW18}.
\subsection{The degenerate case $n$=1}
In this particular case, the gravitational part of the action Eq.~\eqref{1} takes the following form~\cite{CCCQG2021}
\begin{equation}\label{250}
S[g]=\frac{1}{2\kappa^2}\int d^4x \sqrt{-g} \left(R+a_{1}R\Box^{-1}R\right)\ ,
\end{equation}
The gravitational wave can be derived as solutions of the linearized field equations of the free action Eq.~\eqref{250}. Thus, according to the system Eqs.~\eqref{199} and \eqref{linearizedsistem}, we get
\begin{equation}\label{250.1}
\begin{gathered}
\Box \bar{h}_{\mu\nu}=0\ ,\\
\left(3a_{1}-B\right)\Box\delta\phi_{1}+3\Box\delta\lambda_{1}=0\ ,\\
a_{1}\Box\delta\phi_{1}-\Box\delta\lambda_{1}=0\ ,
\end{gathered} 
\end{equation}
that in momentum space, according to decomposition Eq.\eqref{200},  equations~\eqref{250.1} yield
\begin{equation}\label{251}
\begin{gathered}
k^{2}\tilde{\bar{h}}_{\mu\nu}=0\ ,\\
k^{2}\left[\left(3a_{1}-B\right)\widetilde{\delta\phi}_{1}+3\widetilde{\delta\lambda}_{1}\right]=0\ ,\\
k^{2}\Bigl(a_{1}\widetilde{\delta\phi}_{1}-\widetilde{\delta\lambda}_{1}\Bigr)=0\ .
\end{gathered} 
\end{equation}
When $k^{2}=0$ Eqs.~\eqref{251} admit the standard plus and times solutions of general relativity. While, when $k^{2}\neq 0$ the following coefficient matrix, that is
\begin{equation}\label{252}
A_{2}=
\begin{pmatrix}
3a_{1}-B	&	3\\
a_{1}			&	-1
\end{pmatrix}\ ,
\end{equation}
has the following determinant 
\begin{equation}\label{253}
\vert A_{2}\vert=(-1)(6a_{1}-B)\ .
\end{equation}
It vanish only when $B=6a_{1}$.  Then, for $B\neq 6a_{1}$ there are no $k^{2}\neq 0$ solutions,  while for  $B=6a_{1}$ we get
\begin{equation}\label{254}
\tilde{\bar{h}}_{\mu\nu}=0 \quad\text{and}\quad\widetilde{\delta\lambda}_{1}=a_{1}\widetilde{\delta\phi}_{1}\ .
\end{equation}
From the gauge Eqs.~\eqref{100},\eqref{101} and \eqref{102} and from relations Eqs.~\eqref{254}, it gets
\begin{equation}\label{255}
\tilde{h}=-\frac{4}{3}\widetilde{\delta\phi}_{1}=\tilde{A}\ ,
\end{equation}
\begin{equation}\label{256}
\tilde{h}_{\mu\nu}=-\frac{1}{3}\eta_{\mu\nu}\widetilde{\delta\phi}_{1}\ ,
\end{equation}
from which the gravitational wave is obtained 
\begin{multline}\label{260}
h_{\mu\nu}\left(x\right)=\frac{1}{\left(2\pi\right)^{3/2}}\int d^{3}\mathbf{k}_{1}\,\widetilde{C}_{\mu\nu}\left(\mathbf{k}_{1}\right)e^{i k_{1}\cdot x}\\+\frac{1}{\left(2\pi\right)^{3/2}}\int d^{3}\mathbf{k}\,\frac{\eta_{\mu\nu}}{4}\tilde{A}\left(\mathbf{k}\right)e^{i k\cdot x}+c.c. \ .
\end{multline}
That is, when $B=6a_{1}$ it can be regarded as a degenerate case, because every value $k^{2}\in \mathbb{R}_{+}$ is solutions of system Eq.~\eqref{251}, given that it is not the solution of any algebraic equation.  If we can set $\phi_{1}^{(0)}=\lambda_{1}^{(0)}=0$, the constraint Eq.~\eqref{constraintBeq0} becomes
\begin{equation}\label{constraintBeq1}
a_{1}=\frac{1}{6}\ ,
\end{equation}
that is, a dimensionless constant because also $\Box^{-1}R$ is dimensionless.The gravitational action Eq.~\eqref{1} takes the simple form
\begin{equation}\label{degenerateaction}
S_{g}[g]=\frac{1}{2\kappa^2}\int d^4x \sqrt{-g} \biggl(R+\frac{1}{6}R\,\Box^{-1}R\biggr)\ ,
\end{equation}
and in units $G=c=1$ in two-dimensional curved spacetime, its non-local part reproduces the Polyakov effective action~\cite{BOOS, AC}
\begin{equation}\label{265}
S_{g}[g]=\frac{1}{96}\int d^2x \sqrt{-g} R\,\Box^{-1}R\ .
\end{equation}
\subsection{The case $n$=2}
In this second case, the gravitational action Eq.~\eqref{1} takes the following form
\begin{equation}\label{270}
S[g]=\frac{1}{2\kappa^2}\int d^4x \sqrt{-g} \left(R+a_{1}R\Box^{-1}R+a_{2}R\Box^{-2}R\right)\ .
\end{equation} 
The differential system in linear approximation Eq.~\eqref{linearizedsistem} becomes 
\begin{equation}\label{271}
\begin{gathered}
\Box \bar{h}_{\mu\nu}=0\ ,\\
\left(3a_{1}-B\right)\Box\delta\phi_{1}++3a_{2}\Box\delta\phi_{2}+3\Box\delta\lambda_{1}=0\ ,\\
\Box\delta\phi_{2}-\delta\phi_{1}=0\ ,\\
a_{1}\Box\delta\phi_{1}-\Box\delta\lambda_{1}+\delta\lambda_{2}=0\ ,\\
a_{2}\Box\delta\phi_{1}-\Box\delta\lambda_{2}=0\ ,
\end{gathered} 
\end{equation}
that in momentum space, according to decomposition Eq.~\eqref{200}, yields
\begin{equation}\label{272}
\begin{gathered}
k^{2}\tilde{\bar{h}}_{\mu\nu}=0\ ,\\
k^{2}\left[\left(3a_{1}-B\right)\widetilde{\delta\phi}_{1}+3a_{2}\widetilde{\delta\phi}_{2}+3\widetilde{\delta\lambda}_{1}\right]=0\ ,\\
\widetilde{\delta\phi}_{1}+k^{2}\widetilde{\delta\phi}_{2}=0\ ,\\
a_{1}k^{2}\widetilde{\delta\phi}_{1}-k^{2}\widetilde{\delta\lambda}_{1}-\widetilde{\delta\lambda}_{2}=0\ ,\\
k^{2}\Bigl(a_{2}\widetilde{\delta\phi}_{1}-\widetilde{\delta\lambda}_{2}\Bigr)=0\ .
\end{gathered} 
\end{equation}
For $k^{2}=0$ the Eq.~\eqref{272} gives back the two standard plus and times gravitational wave,  proper to general relativity.  Now let's see under what conditions non-null waves like $k^{2}\neq 0$ are obtained. 
Then, the coefficient matrix of the last four equations of the Eq.~\eqref{272} yields
\begin{equation}\label{275}
A_{4}=
\begin{pmatrix}
3a_{1}-B	&	3a_{2}	&		3		&		0\\
1				&	k^{2}	&		0		&		0\\
a_{1}k^{2}	&	0			&	-k^{2}	&		-1\\
a_{2}		&	0			&		0		&		-1
\end{pmatrix}\ ,
\end{equation}
whose determinant takes a form as 
\begin{equation}\label{276}
\vert A_{4}\vert=k^{2}\left[(6a_{1}-B)k^{2}-6a_{2}\right]\ .
\end{equation}
The first equation of Eq.~\eqref{272} gives $\bar{h}_{\mu\nu}=0$ and its trace vanishes $\bar{h}=0$. Thus, if $B=6a_{1}$, there is no non-null solutions because $\vert A_{4}\vert\neq 0$, while for $B\neq 6a_{1}$ the determinant vanishes for one non-null solution
\begin{equation}\label{277}
k_{2}^{2}=\frac{6a_{2}}{6a_{1}-B}\ ,
\end{equation}
if 
\begin{equation}\label{278}
\frac{6a_{2}}{6a_{1}-B}> 0\ .
\end{equation}
Hence from Eq.~\eqref{209}, we have
\begin{equation}\label{279}
\tilde{h}_{\mu\nu}(\mathbf{k}_{2})=-\frac{1}{3}\eta_{\mu\nu}\tilde{\phi}_{1}(\mathbf{k}_{2})=\frac{\eta_{\mu\nu}}{4}\tilde{A}_{2}(\mathbf{k}_{2})\ .
\end{equation}
Then, the combination of null and non-null waves $k_{1}$ and $k_{2}$ give us
\begin{multline}\label{279}
h_{\mu\nu}\left(x\right)=\frac{1}{\left(2\pi\right)^{3/2}}\int d^{3}\mathbf{k}\,\widetilde{C}_{\mu\nu}\left(\mathbf{k}\right)e^{i k_{1}\cdot x}\\+\frac{1}{\left(2\pi\right)^{3/2}}\int d^{3}\mathbf{k}\,\frac{\eta_{\mu\nu}}{4}\tilde{A}_{2}\left(\mathbf{k}\right)e^{i k_{2}\cdot x}+c.c.\ .
\end{multline}
\section{Polarizations via geodesic deviation equation and helicity}\label{sec4}
The equation of geodesic deviation provides an effective tool to investigate the polarizations of gravitational radiation when the wave invests a small region of spacetime, as a measure of relative acceleration between nearby geodesic, on which two freely test masses move slowly.  Hence, we consider a wave $h_{\mu\nu}\left(t-v_{g}z\right)$  propagating in  $+\hat{z}$ direction, where $v_{g}$ is the group velocity in units where $c=1$ defined as
\begin{equation}\label{300}
v_{g}=\frac{d\omega}{d k_{z}}\ ,
\end{equation} 
that for the dispersion relation of ours $j$-th modes Eq.~\eqref{210.2} becomes 
\begin{equation}\label{300.1}
v_{g_{j}}=\frac{c^{2}k_{z}}{\sqrt{k_{z}^{2}+M_{j}^{2}}}\ .
\end{equation}
Now, let us start from the geodesic deviation equation for only spatial components in a locally Lorentz normal coordinate system, that reduces to the form 
\begin{equation}\label{eqdevgeoelectric}
\ddot x^{i}=-R^{i}_{\phantom{i}0k0}x^{k}\ ,
\end{equation}
with the Latin index range over the set $\left\{1,2,3\right\}$. The separation vector is $x^{\mu}$ and $R^{i}_{\phantom{i}0k0}$ are the only six measurable components, so-called electric components~\cite{STRAGR}.  Inserting the linearized electric components of the Riemann tensor $R^{\left(1\right)}_{\phantom{1}i0j0}$ expressed in terms of the metric perturbation $h_{\mu\nu}$ 
\begin{equation}\label{301}
R^{\left(1\right)}_{\phantom{1}i0j0}=\frac{1}{2}\left(h_{i0,j0}+h_{j0,i0}-h_{ij,00}-h_{00,ij}\right)\ ,
\end{equation}
into Eq.~\eqref{eqdevgeoelectric},  we find a linear non-homogeneous system of differential equations  
\begin{equation}\label{eqdevgeolinear}
\begin{cases}
\ddot x(t)=-\frac{1}{2}h_{11,00}\;x-\frac{1}{2}h_{12,00}\;y+\frac{1}{2}\left(h_{10,03}-h_{13,00}\right)z\\
\ddot y(t)=-\frac{1}{2}h_{12,00}\;x-\frac{1}{2}h_{22,00}\;y+\frac{1}{2}\left(h_{02,03}-h_{23,00}\right)z\\
\ddot z(t)=\frac{1}{2}\left(h_{01,03}-h_{13,00}\right)x\\
\qquad\qquad\qquad\quad+\frac{1}{2}\left(h_{02,03}-h_{23,00}\right)y+\frac{1}{2}\left(2h_{03,03}-h_{33,00}-h_{00,33}\right)z
\end{cases}\ .
\end{equation}
Thus,  we keep $k_{1}^{\mu}=\left(\omega_{1},0,0,k_{z}\right)$ four-vector fixed,  which from dispersion relation Eq.~\eqref{204.1}, means to keep $k_{z}$.  For a massless plane wave traveling in  $+\hat{z}$ direction,  that is, $k_{1}^{2}=0$,  which propagates at speed $c$,  Eq.~\eqref{GWsNonLocalVacuum} yields    
\begin{equation}\label{FirstOrderTetradmassless}
h^{(k_{1})}_{\mu\nu}\left(t,z\right)=\sqrt{2}\left[\tilde{\epsilon}^{(+)}\left(\omega_{1}\right)\epsilon^{(+)}_{\mu\nu}+\tilde{\epsilon}^{(\times)}\left(\omega_{1}\right)\epsilon^{(\times)}_{\mu\nu}\right]e^{i\omega_{1}\left(t-z\right)}+c.c.\ ,
\end{equation}
with two polarization tensors 
\begin{equation}\label{302}
\epsilon^{(+)}_{\mu\nu}=\frac{1}{\sqrt{2}}
\begin{pmatrix} 
0 & 0 & 0 & 0 \\
0 & 1 & 0 & 0 \\
0 & 0 & -1 & 0 \\
0 & 0 & 0 & 0
\end{pmatrix}\ ,
\end{equation}
\begin{equation}\label{303}
\epsilon^{(\times)}_{\mu\nu}=\frac{1}{\sqrt{2}}
\begin{pmatrix} 
0 & 0 & 0 & 0 \\
0 & 0 & 1 & 0 \\
0 & 1 & 0 & 0 \\
0 & 0 & 0 & 0
\end{pmatrix}\ ,
\end{equation}
and angular frequency $\omega_{1}=k_{z}$.  Moreover,  when $k_{j}^{2}\neq 0$,i.e., for a massive plane wave, propagating in $+\hat{z}$ direction keeping $k_{j}^{\mu}=\left(\omega_{j},0,0,k_{z}\right)$ fixed, that according to dispersion relation Eq.~\eqref{210.2} means $k_{z}$ fixed,  Eq.~\eqref{GWsNonLocalVacuum} becomes 
\begin{equation}\label{FirstOrderTetradmassive}
h^{(k_{j})}_{\mu\nu}\left(t,z\right)=\frac{\tilde{A}_{j}\left(k_{z}\right)}{4}\eta_{\mu\nu}e^{i\left(\omega_{j}t-k_{z}z\right)}+c.c.\ ,
\end{equation}
where here the propagation speed of $j$-th mode is less than $c$ and $j$ run over $\{2,\dots,n\}$.  In a more compact form,  the metric linear perturbation  $h_{\mu\nu}$,  traveling in the $+\hat{z}$ direction with $k_{z}$ fixed,  is given by
\begin{multline} \label{304}
h_{\mu\nu}\left(t,z\right)=\sqrt{2}\left[\tilde{\epsilon}^{(+)}\left(\omega_{1}\right)\epsilon^{(+)}_{\mu\nu}+\tilde{\epsilon}^{(\times)}\left(\omega_{1}\right)\epsilon^{(\times)}_{\mu\nu}\right]e^{i\omega_{1}\left(t-z\right)}\\
+\sum_{j=2}^{n}\tilde{\epsilon}^{\left(s_{j}\right)}_{\mu\nu}\left(k_{z}\right)e^{i\left(\omega_{j}t-k_{z}z\right)}+c.c.\ ,
\end{multline}
where $\tilde{\epsilon}^{\left(s_{j}\right)}_{\mu\nu}$ is the polarization tensor associated to the mixed scalar mode 
\begin{equation}\label{305}
\tilde{\epsilon}^{\left(s_{j}\right)}_{\mu\nu}\left(k_{z}\right)=\frac{\eta_{\mu\nu}}{4}\tilde{A}_{j}\left(k_{z}\right)=\left(\epsilon^{(TT)}_{\mu\nu}-\sqrt{2}\epsilon_{\mu\nu}^{(b)}-\epsilon_{\mu\nu}^{(l)}\right)\frac{\tilde{A}_{j}\left(k_{z}\right)}{4}\ ,
\end{equation}
and the three polarization tensors are explicitly the following
\begin{align}\label{306}
\epsilon^{(TT)}_{\mu\nu}&=
\begin{pmatrix} 
1 & 0 & 0 & 0 \\
0 & 0 & 0 & 0 \\
0 & 0 & 0 & 0 \\
0 & 0 & 0 & 0
\end{pmatrix}\ , &
\epsilon^{(b)}_{\mu\nu}&=\frac{1}{\sqrt{2}}
\begin{pmatrix} 
0 & 0 & 0 & 0 \\
0 & 1 & 0 & 0 \\
0 & 0 & 1 & 0 \\
0 & 0 & 0 & 0
\end{pmatrix}\ , &
\epsilon^{(l)}_{\mu\nu}&=
\begin{pmatrix} 
0 & 0 & 0 & 0 \\
0 & 0 & 0 & 0 \\
0 & 0 & 0 & 0 \\
0 & 0 & 0 & 1
\end{pmatrix}\ .
\end{align}
The set of polarization tensors $\left\{\epsilon_{\mu\nu}^{\left(+\right)}, \epsilon_{\mu\nu}^{\left(\times\right)}, \epsilon_{\mu\nu}^{\left(TT\right)}, \epsilon_{\mu\nu}^{\left(b\right)}, \epsilon_{\mu\nu}^{\left(l\right)}\right\}$ fulfill the orthonormality relations
\begin{equation}\label{306}
\text{Tr}\left\{\epsilon^{(a)}\epsilon^{(b)}\right\}=\epsilon_{\mu\nu}^{(a)}\epsilon^{(b)\mu\nu}=\delta^{ab}\quad\text{with}\quad a,b\in\left\{+, \times, TT, b, l\right\}\ .
\end{equation} 
The helicity of wave Eq.~\eqref{304} can be derived by performing a rotation of $\theta$ angle along the $z$ propagation direction, on the polarization tensors as follows
\begin{equation}
\epsilon_{\mu\nu}^{\prime}=R_{\mu}^{\alpha}\epsilon_{\alpha\beta}R_{\beta}^{\nu}\ ,
\end{equation}
where the matrix of rotation is
\begin{equation}
R=
\begin{pmatrix}
1	&	0	&	0	&	0\\
0	&	1	&	0	&	0\\
0	&	0	&	\cos\theta	&	\sin\theta	\\
0	&	0	&	-\sin\theta	&	\cos\theta	
\end{pmatrix}\ .
\end{equation}
Hence the wave has helicity $s$ is if polarization tensor changes as
\begin{equation}
\epsilon_{\mu\nu}^{\prime}=e^{\pm i s\theta}\epsilon_{\mu\nu}\ .
\end{equation}
Thus, for the $\omega_{1}$ mode with $k_{1}^{2}=0$, the polarization tensors $\epsilon_{\mu\nu}^{(+)}$ and $\epsilon_{\mu\nu}^{(\times)}$ transform as
\begin{equation}
\epsilon^{\prime}=e^{\pm 2i\theta}\epsilon\ ,
\end{equation}
that is, we obtain the tensor modes with helicity equal to two. While for $\omega_{j}$ mode with $k_{j}^{2}\neq 0$, the polarization tensors $\epsilon_{\mu\nu}^{(s_{j})}$ transform as 
\begin{equation}
\Bigl(\epsilon^{(s_{j})}\Bigr)^{\prime}=\epsilon^{(s_{j})}\ ,
\end{equation}
that is, we obtain $j$ scalar modes with helicity equal to zero. The polarization tensor of the single scalar $j$-th mode $\tilde{\epsilon}^{\left(s_{j}\right)}_{\mu\nu}$, is a mixed state made up of a combination of longitudinal and transverse scalar modes,  produced by the single degree of freedom $\tilde{A}_{j}$, similarly to $f(R)$ gravity which has three d.o.f.: $\tilde{\epsilon}^{(+)}$, $\tilde{\epsilon}^{(\times)}$ and $\tilde{A}$~\cite{IKEDA}.  However, as we will see, the transverse component weighs more than the longitudinal one if we consider slightly massive waves.  Then, considering that the polarization of the wave is determined only by the spatial components of the polarization tensor $\tilde{\epsilon}^{\left(s_{j}\right)}_{\mu\nu}$,  we use its restriction to only spatial components, i.e., $\tilde{\epsilon}^{\left(s_{j}\right)}_{m,n}$, that give
\begin{equation}\label{307}
\tilde{\epsilon}^{\left(s_{j}\right)}_{m,n}=\Bigl(\sqrt{2}\epsilon_{m,n}^{(b)}+\epsilon_{m,n}^{(l)}\Bigl)\frac{\tilde{A}_{j}\left(k_{z}\right)}{4}\ ,
\end{equation}
where $(m,n)$ range over $(1,2,3)$.  
Thus,  inserting the generic gravitational wave $z$-axes propagating at $k_{z}$ fixed into the system Eq.~\eqref{eqdevgeolinear},  we get in geometrized units where $c=G=1$
\begin{equation}\label{307.1}
\left\{
\begin{array}{l}
\ddot x(t)=\frac{1}{2}\omega_{1}^{2}\left[\tilde{\epsilon}^{\left(+\right)}\left(\omega_{1}\right)x+\tilde{\epsilon}^{\left(\times\right)}\left(\omega_{1}\right)y\right]e^{i\omega_{1}\left(t-z\right)}\\
\qquad\qquad\qquad\qquad\qquad\qquad-\frac{1}{8}\sum_{j=2}^{n}\omega^{2}_{j}\tilde{A}_{j}\left(k_{z}\right)x e^{i\left(\omega_{j}t-k_{z}z\right)}+c.c.\\ \\
\ddot y(t)=\frac{1}{2}\omega_{1}^{2}\left[\tilde{\epsilon}^{\left(\times\right)}\left(\omega_{1}\right)x-\tilde{\epsilon}^{\left(+\right)}\left(\omega_{1}\right)y\right]e^{i\omega_{1}\left(t-z\right)}\\
\qquad\qquad\qquad\qquad\qquad\qquad-\frac{1}{8}\sum_{j=2}^{n}\omega^{2}_{j}\tilde{A}_{j}\left(k_{z}\right)y e^{i\left(\omega_{j}t-k_{z}z\right)}+c.c.\\ \\
\ddot z(t)=-\frac{1}{8}\sum_{j=2}^{n}M^{2}_{j}\tilde{A}_{j}\left(k_{z}\right)ze^{i\left(\omega_{j}t-k_{z}z\right)}+c.c.
\end{array}
\right.\ ,
\end{equation}
that can be integrated, assuming that displacements are small,  as 
\begin{equation}\label{307.2}
\begin{cases}
x(t)=x_{0}+\frac{1}{2}\left[\tilde{\epsilon}^{\left(+\right)}x_{0}+\tilde{\epsilon}^{\left(\times\right)}y_{0}\right]e^{i\omega_{1}\left(t-z\right)}\\
\qquad\qquad\qquad\qquad+\frac{1}{8}\sum_{j=2}^{n}\tilde{A}_{j}\left(k_{z}\right)x_{0}e^{i\left(\omega_{j}t-k_{z}z\right)}+c.c.\\ \\
y(t)=y_{0}+\frac{1}{2}\left[\tilde{\epsilon}^{\left(\times\right)}x_{0}-\tilde{\epsilon}^{\left(+\right)}y_{0}\right]e^{i\omega_{1}\left(t-z\right)}\\
\qquad\qquad\qquad\qquad+\frac{1}{8}\sum_{j=2}^{n}\tilde{A}_{j}\left(k_{z}\right)y_{0}e^{i\left(\omega_{j}t-k_{z}z\right)}+c.c.\\ \\
z(t)=z_{0}+\frac{1}{8\omega^{2}_{j}}M^{2}_{j}\tilde{A}_{j}\left(k_{z}\right)z_{0}e^{i\left(\omega_{j}t-k_{z}z\right)}+c.c.
\end{cases}.
\end{equation}
Hence,  the solution Eqs.~\eqref{307.2} in the  case $k_{1}^{2}=0$ for massless waves $h_{\mu\nu}^{(k_{1})}$,  that is, exactly null plane, associated to angular frequency $\omega_{1}$,  gives
\begin{equation}\label{308}
\begin{cases}
x(t)=x_{0}+\frac{1}{2}\left[\tilde{\epsilon}^{\left(+\right)}x_{0}+\tilde{\epsilon}^{\left(\times\right)}y_{0}\right]e^{i\omega_{1}\left(t-z\right)}+c.c.\\ 
y(t)=y_{0}+\frac{1}{2}\left[\tilde{\epsilon}^{\left(\times\right)}x_{0}-\tilde{\epsilon}^{\left(+\right)}y_{0}\right]e^{i\omega_{1}\left(t-z\right)}+c.c.\\ 
z(t)=z_{0}+c.c.
\end{cases}\ ,
\end{equation}
namely, we recover the two pure,  plus and cross, standard transverse tensor modes predicted by general relativity,.  

Otherwise,  in the pure case of a massive wave $h^{(k_{j})}_{\mu\nu}$, i.e, non-null plane waves, with $0\neq k_{j}^{2}=M^2_{j}=\omega^{2}_{j}-k^{2}_{z}$ for the $j$-th mode of angular frequency $\omega_{j}$,  the solution  Eq.~\eqref{307.2} becomes
\begin{equation}\label{solutionsldgs}
\left\{
\begin{array}{l}
x(t)=x(0)+\frac{1}{8}\tilde{A}_{j}\left(k_{z}\right)x(0)e^{i\left(\omega_{j}t-k_{z}z\right)}+c.c.\\ \\
y(t)=y(0)+\frac{1}{8}\tilde{A}_{j}\left(k_{z}\right)y(0)e^{i\left(\omega_{j}t-k_{z}z\right)}+c.c.\\ \\
z(t)=z(0)+\frac{1}{8\omega^{2}_{j}}M^{2}_{j}\tilde{A}_{j}\left(k_{z}\right)z(0)e^{i\left(\omega_{j}t-k_{z}z\right)}+c.c.
\end{array}.
\right.
\end{equation}
When we are sufficiently far from the  radiation source, we can suppose that $M^2_{j}$ is very small, i.e., that the plane waves are nearly null.  Then, we can expand our quantities with respect to a parameter $\gamma$, which takes into account the difference in speed between nearly null plane waves with speed $v_{g}$ and exactly null plane waves with speed $c$ where the parameter $\gamma$ vanishes. Also, the expansion coefficient $\gamma$ reads~\cite{CMW} 
\begin{equation}\label{310}
\gamma=\left(\frac{c}{v_{g}}\right)^{2}-1\ .
\end{equation} 
By using Landau symbols,  namely little-$o$ and big-$\mathcal{O}$ notation,  we can keep $k_{z}$ fixed and expand in terms of our parameter $\gamma$ the following quantities  
\begin{equation}\label{311}
\frac{M^{2}_{j}}{k_{z}^{2}}=\frac{1}{k_{z}^{2}}\left[\frac{\omega_{j}^{2}}{c^{2}}-k_{z}^{2}\right]=\left(\frac{\omega_{j}}{c k_{z}}\right)^{2}-1=\gamma\ ,
\end{equation}
and in $c=1$ units we have 
\begin{equation}\label{312}
\frac{\omega_{j}}{k_{z}}=\sqrt{1+\gamma}=\left(1+\frac{1}{2}\gamma\right)+o\left(\gamma\right)\ ,
\end{equation}
\begin{equation}\label{313}
e^{i\left(\omega_{j}t-k_{z}z\right)}=e^{ik_{z}\left(t-z\right)}+\mathcal{O}\left(\gamma\right)\ ,
\end{equation}
from which it gets to first order in $\gamma$
\begin{equation}\label{314}
\frac{M^{2}_{j}}{\omega_{j}^{2}}=\gamma+o\left(\gamma\right)\ .
\end{equation}
Hence, taking into account the previous expansions, the solution Eq.~\eqref{solutionsldgs} for the mode $\omega_{j}$ give us 
\begin{equation}\label{315}	
\left\{
\begin{array}{l}
x(t)=x(0)+\frac{1}{8}\tilde{A}_{j}\left(k_{z}\right)x(0)e^{i k_{z}\left(t-z\right)}+\mathcal{O}\left(\gamma\right)+c.c.\\ \\
y(t)=y(0)+\frac{1}{8}\tilde{A}_{j}\left(k_{z}\right)y(0)e^{i k_{z}\left(t-z\right)}+\mathcal{O}\left(\gamma\right)+c.c.\\ \\
z(t)=z(0)+\frac{1}{8}\gamma\tilde{A}_{j}\left(k_{z}\right)z(0)e^{i k_{z}\left(t-z\right)}+\mathcal{O}\left(\gamma^{2}\right)+c.c.
\end{array}\ ,
\right.
\end{equation}
that is 
\begin{equation}\label{316}
\Delta z(t)=o\left(\Delta x(t)\right)\quad\text{and}\quad\Delta z(t)=o\left(\Delta y(t)\right)\quad \text{for}\quad\gamma\to 0\ .
\end{equation}
This suggests that longitudinal modes are infinitesimal in higher order than transverse modes when $\gamma$ tends to zero and so only the breathing tensor polarization $\epsilon^{(b)}_{m,n}$ survives. Thus, when a gravitational wave strikes a sphere of particles of radius $r=\sqrt{x^2(0)+y^{2}(0)+z^{2}(0)}$, from the Eq.~\eqref{315} the radiation will distort it into an ellipsoid described by
\begin{equation}\label{307}
\left(\frac{x}{\rho_{1}(t)}\right)^{2}+\left(\frac{y}{\rho_{1}(t)}\right)^{2}+\left(\frac{z}{\rho_{2}(t)}\right)^{2}=r^{2}\ ,
\end{equation}
where the principal axes to zero-th order in $\gamma$ are given by
\begin{gather}
\rho_{1}(t)=1+\frac{1}{4}\tilde{A}_{j}\left(k_{z}\right)\cos\left[k_{z}\left(t-z\right)+\phi\right]+\mathcal{O}\left(\gamma\right)\ ,\\
\rho_{2}(t)=1+\mathcal{O}\left(\gamma\right)\ .
\end{gather} 
So, to lowest order only $\rho_{1}$ varies between its maximum and minimum value.  This means that the ellipsoid swings only on $xy$-plane between two circumferences of minimum and maximum radius, reproducing the additional transverse scalar breathing polarization which has zero helicity to lowest order in $\gamma$~\cite{PW}. 

According to these considerations,  under constraint $B\neq 6a_{1}$,  the gravity $R+\sum_{h=1}^{n}a_{h}R\Box^{-h}R$,  linear in $R$ and $\Box^{-k}$~\cite{CCN},  has $n+1$ degrees of freedom: two of these,  $\tilde{\epsilon}^{\left(+\right)}$ and $\tilde{\epsilon}^{\left(\times\right)}$, generate the standard  tensor modes of general relativity, while each $n-1$ degrees of freedom $\tilde{A}_{j}$ gives rise to a further breathing scalar mode.  Instead, for $B=6a_{1}$ one scalar mode is inhibited for all $n$ greater than two, while in $n=1$ a degenerate state with continuous infinity of solutions $k^{2}\neq 0$ occurs.  In summary,   generally $R+\sum_{h=1}^{n}a_{h}R\Box^{-h}R$ gravity has three polarizations, $(+)$, $(\times)$ and $(b)$, but $n+1$ modes,  namely two massless 2-helicity  transverse tensor modes and $n-1$ massive 0-helicity scalar mode,  all purely transverse to lowest order in $\gamma$,  exactly like $f(R)$ gravity (see for a discussion~\cite{IKEDA, AMA, BCLN, CCL, CCDLV}). 
\section{Polarizations via NP formalism and helicity}\label{sec5}
Newman-Penrose (NP) formalism is a additional method to analyze polarizations that works out for massless waves.  Even if it is not directly applicable to massive waves,  NP formalism can be applied to slightly massive gravitational waves generalizing it to waves  propagating along nearly null geodesics~\cite{CMW,NP}.  Furthermore, the little group $E\left(2\right)$ classification fails for massive waves but can be recovered to first order in the small expansion parameter $\gamma$, that is for slightly massive waves. Let us define a local quasi-orthonormal null tetrad basis $\left(k, l, m, \bar{m}\right)$,  as a new basis~\cite{PW, ELL, ELLWW}
\begin{align}
k&=\partial_{t}+\partial_{z}\ ,
& l&=\frac{1}{2}\left(\partial_{t}-\partial_{z}\right)\ ,\label{318.1}\\
m&=\frac{1}{\sqrt{2}}\left(\partial_{x}+i\partial_{y}\right)\ ,
& \bar{m}&=\frac{1}{\sqrt{2}}\left(\partial_{x}-i\partial_{y}\right)\ ,\label{318.2}
\end{align}
which has to fulfill the relations
\begin{gather}
k\cdot l=-m\cdot\bar{m}=1\ , \nonumber\\
k\cdot k=l\cdot l=m\cdot m=\bar{m}\cdot\bar{m}=0\ ,\label{319}\\
k\cdot m=k\cdot\bar{m}=l\cdot m=l\cdot\bar{m}=0\ ,\nonumber
\end{gather}
that is, the Minkowski metric tensor $\eta_{\mu\nu}$ of signature $-2$,  can be expressed as 
\begin{equation}\label{320}
\eta^{\mu\nu}=2k^{(\mu}l^{\nu)}-2m^{(\mu}\bar{m}^{\nu)}\ .
\end{equation}
Therefore we can raise and lower the tetrad indices by the metric of the tetrad $\eta_{ab}$
\begin{equation}\label{321}
\eta_{ab}=\eta^{ab}=
\begin{pmatrix}
0	&	1	&	0	&	0\\
1	&	0	&	0	&	0\\
0	&	0	&	0	&	-1\\
0	&	0	&	-1	&	0
\end{pmatrix} \ ,
\end{equation}
where $(a,b)$ run over $(k, l, m, \bar{m})$. We now can decompose the Riemann tensor into three irreducible parts, namely Ricci decomposition, related to Weyl tensor,  Ricci tensor and Ricci scalar. The four-dimensional Weyl tensor $C_{\mu\nu\rho\sigma}$ in coordinate basis, is defined as the trace free part of the Riemann tensor 
\begin{equation}\label{322}
C_{\mu\nu\rho\sigma}=R_{\mu\nu\rho\sigma}-2g_{[\mu|[\rho}R_{\sigma]|\nu]}+\frac{1}{3}g_{\mu[\rho}g_{\sigma]\nu}R\ ,
\end{equation}
that in tetrad form becomes
 \begin{equation}\label{TetradWeylTensor}
 C_{abcd}=R_{abcd}-2\eta_{[a|[c}R_{d]|b]}+\frac{1}{3}\eta_{a[c}\eta_{d]b}R\ ,
 \end{equation}
according to tetrad components of the generic tensor $P_{a b c d \dots}$ expressed in terms of the local coordinate basis as
\begin{equation}\label{323}
P_{a b c d \cdots}=P_{\alpha,\beta\gamma\delta\cdots}a^{\alpha}b^{\beta}c^{\gamma}d^{\delta}\cdots\ ,
\end{equation}
where $(a, b, c, d, \dots)$ run over $(k, l, m, \bar{m})$.  So, can be defined fifteen Newman-Penrose quantities,  expressing the Weyl tensor, the Ricci tensor and the Ricci scalar in the null tetrad basis Eqs.~\eqref{318.1} and \eqref{318.2}. The NP~amplitudes are, specifically,  the five  Weyl-NP $\Psi$  complex scalars $\{\Psi_{0},\Psi_{1},\Psi_{2},\Psi_{2},\Psi_{4}\}$, expressed in tetrad components of the Weyl tensor as
\begin{equation}\label{323.1}
\begin{aligned}
 \Psi_{0}&\equiv -C_{kmkm}\ ,\\
\Psi_{1}&\equiv -C_{klkm}\ ,\\
 \Psi_{2}&\equiv -C_{km\bar{m}l}\ ,\\
 \Psi_{3}&\equiv -C_{kl\bar{m}l}\ ,\\
 \Psi_{4}&\equiv -C_{\bar{m}l\bar{m}l}\ ,
\end{aligned}
\end{equation}
 and the seven Ricci-NP scalars $\Phi$ and $\Lambda$,  of which four real $\{\Phi_{00},\Phi_{11},\Phi_{22},\Lambda\}$ and three complex $\{\Phi_{10},\Phi_{20},\Phi_{21}\}$, expressed in tetrad components of  Ricci tensor  and Ricci scalar as 
\begin{equation}\label{324}
\begin{aligned}
&\quad\Phi_{02}\equiv-\frac{1}{2}R_{mm\ ,}\\
&\left\{
\begin{array}{l}
\Phi_{01}\equiv-\frac{1}{2}R_{km}\\
\Phi_{12}\equiv-\frac{1}{2}R_{lm}\\
\end{array}\ ,
\right.\\
&\left\{
\begin{array}{l}
\Phi_{00}\equiv-\frac{1}{2}R_{kk}\\
\Phi_{11}\equiv-\frac{1}{4}\left(R_{kl}+R_{m\bar{m}}\right)\\
\Phi_{22}\equiv-\frac{1}{2}R_{ll}
\end{array}\ ,
\right. \\
&\left\{
\begin{array}{l}
\Phi_{10}\equiv-\frac{1}{2}R_{k\bar{m}}=\Phi_{01}^{*}\\
\Phi_{21}\equiv-\frac{1}{2}R_{l\bar{m}}=\Phi_{21}^{*}\\
\end{array}\ ,
\right.\\
&\quad\Phi_{20}\equiv-\frac{1}{2}R_{\bar{m}\bar{m}}=\Phi_{02}^{*}\ ,\\
&\qquad\Lambda\equiv\frac{R}{24}\ .
\end{aligned}
\end{equation}
For the purpose of expand the slightly massive gravitational radiation in terms of null plane waves, we define the four-wavevector $\left(k_{2}^{\prime}\right)^{\mu}$ in units $c=1$ associated to the nearly null plane wave traveling in $+z$~direction as 
\begin{equation}\label{325}
k_{j}^{\mu}=\left(\omega_{j}, 0, 0, k_{z}\right)=\omega_{j}k_{j}^{\prime\mu}\ ,
\end{equation}
with $j\in\{2,\ldots,n\}$ where
\begin{equation}\label{326}
k^{\prime\mu}_{j}=\left(1, 0, 0, v_{g_{j}}\right)\ .
\end{equation}
Now, we set the retarded time $\tilde{u}$ as
\begin{equation}\label{327}
\tilde{u}=t-v_{g}z\ ,
\end{equation}
that gives us
\begin{equation}\label{328}
\nabla_{\mu}\tilde{u}=\left(k_{j}^{\prime}\right)_{\mu}\ .
\end{equation}
So, we expand $k_{j}^{\prime}$ with respect to  NP~tetrads basis as
\begin{equation}\label{kprimeresklm}
k_{j}^{\prime\mu}=\left(1+\gamma_{k}\right)k^{\mu}+\gamma_{l}l^{\mu}+\gamma_{m}m^{\mu}+\bar{\gamma}_{m}\bar{m}^{\mu}\ ,
\end{equation} 
where the expansion coefficients $\gamma_{k}, \gamma_{l}, \gamma_{m}$ are of same order of the $\gamma$, already previous defined in Eq.~\eqref{310}.  Given the freedom of the observer to orient his reference system,  it is possible to choose orientation in such a way that  $k_{j}^{0}=k^{0}$ and $k_{j}^{3}\propto k^{3}$,  where $k^{0}$ is the angular frequency of null wave and  $k^{3}$ the third component of its vector wave. Hence,  Eq.~\eqref{kprimeresklm} gives $\gamma_{l}=-2\gamma_{k}$ and $\gamma_{m}=0$,  and this implies 
\begin{equation}\label{329}
k_{j}^{\prime\mu}=k^{\mu}+\gamma_{k}\left(k^{\mu}-2l^{\mu}\right)\ ,
\end{equation}
or 
\begin{equation}\label{330}
k_{j }^{\prime\mu}=\left(1, 0, 0, 1+2\gamma_{k}\right)\ .
\end{equation}
Then, we note that the parameters $\gamma$ and $\gamma_{k}$ are of the same order, according to 
\begin{equation}\label{331}
\gamma_{k}=-\frac{1}{4}\gamma+\small{o}\left(\gamma\right).
\end{equation}
The derivatives of Riemann tensor can be expressed as
\begin{equation}\label{332}
R_{\alpha\beta\gamma\delta,\mu}=\frac{\partial R_{\alpha\beta\gamma\delta}}{\partial\tilde{u}}\nabla_{\mu}\tilde{u}=\left(k_{j}^{\prime}\right)_{\mu}\dot R_{\alpha\beta\gamma\delta}\ ,
\end{equation}
where the superscripted dot means the derivative with respect to $\tilde{u}$.  From  following identities 
\begin{align}
R_{\alpha\beta\gamma\delta,k}&=-2\gamma_{k}\dot R_{\alpha\beta\gamma\delta}\ ,\label{333}\\
R_{\alpha\beta\gamma\delta,l}&=\left(1+\gamma_{k}\right)\dot R_{\alpha\beta\gamma\delta}\ ,\label{334}\\
R_{\alpha\beta\gamma\delta,m}&=0\ ,\label{335}
\end{align}
joined with differential Bianchi identity 
\begin{equation}\label{336}
R_{ab[cd,e]}=0\ ,
\end{equation}
to zeroth order in  $\gamma_{k}$, the only non-zero tetrad components of Riemann tensor $R_{abcd}$ that remain, are terms of form $R_{lplq}$ with $(p,q)$ range over $(k, m, \bar{m)}$.  In the nearly null plane waves approximation,  only four complex NP~tetrad components are independent and non-vanishing to first order in $\gamma_{k}$, that is, from Eq.~\eqref{TetradWeylTensor}, they are
\begin{align}
\Psi_{2}\left(\tilde{u}\right)&=-\frac{1}{6}R_{lklk}+\mathcal{O}\left(\gamma_{k}\right)\ ,\label{337}\\
\Psi_{3}\left(\tilde{u}\right)&=-\frac{1}{2}R_{lkl\bar{m}}+\mathcal{O}\left(\gamma_{k}\right)\ ,\label{338}\\
\Psi_{4}\left(\tilde{u}\right)&=-R_{l\bar{m}l\bar{m}}\ ,\label{339}\\
\Phi_{22}\left(\tilde{u}\right)&=-R_{lml\bar{m}}\ .\label{340}
\end{align} 
The four amplitudes $\{\Psi_{2}, \Psi_{3}, \Psi_{4}, \Phi_{22}\}$ expressed in terms of tetrad components of metric perturbation $h_{ab}$, are given by 
\begin{align}
\Psi_{2}\left(\tilde{u}\right)&=\frac{1}{12}\ddot{h}_{kk}+\mathcal{O}\left(\gamma_{k}\right)\ ,\label{341}\\
\Psi_{3}\left(\tilde{u}\right)&=\frac{1}{4}\ddot{h}_{k\bar{m}}+\mathcal{O}\left(\gamma_{k}\right)\ ,\label{342}\\
\Psi_{4}\left(\tilde{u}\right)&=\frac{1}{2}\ddot{h}_{\bar{m}\bar{m}}+\mathcal{O}\left(\gamma_{k}\right)\ ,\label{343}\\
\Phi_{22}\left(\tilde{u}\right)&=\frac{1}{2}\ddot{h}_{m\bar{m}}+\mathcal{O}\left(\gamma_{k}\right)\ .\label{344}
\end{align}
These NP~scalars, according to their behavior under the subgroup of Lorentz transformations which leaves  $\mathbf{k}_{j}$ unchanged, namely the little group $E(2)$, show the following four helicity values $s$, for each of them
\begin{align}
\Psi_{2}\quad s=0\ ,& &\Psi_{4\quad}  s=2\ ,&\label{345}\\
\Psi_{3}\quad  s=1\ ,& &\Phi_{22}\quad  s=0\ .&\label{346}
\end{align}
These NP~quantities allow the classification of quasi-Lorentz invariant  GWs, know as Petrov classification.  For a gravitational radiation propagating along  $+\hat{z}$ axis, we can express the four NP~amplitudes both in terms of the electric components of the Riemann tensor $R_{i0j0}$ and its linearized components from Eq.~\eqref{301}. Indeed,  considering the identities 
\begin{align}
R_{kml\bar{m}}&=0&\rightarrow &\,\,\,R_{1313}+R_{2323}=R_{0101}+R_{0202}\ ,\label{347}\\
R_{kmk\bar{m}}&=0&\rightarrow &\,\,\,-2R_{0131}-2R_{0232}=2\left(R_{0101}+R_{0202}\right)\ ,\label{348}\\
R_{klkm}&=0&\rightarrow&\,\,\,R_{0301}=-R_{0331}\quad\text{and}\quad R_{0302}=-R_{0332}\ ,\label{349}\\
R_{kmlm}&=0&\rightarrow &\,\,\,R_{3132}=R_{0102}\quad\text{and}\quad R_{3131}-R_{3232}=R_{0101}-R_{0202}\ ,\label{350}\\
R_{kmkm}&=0&\rightarrow
&\,\,\,\begin{cases}\label{351}
2R_{3101}-2R_{3202}=-R_{3131}+R_{3232}-R_{0101}+R_{0202}\\
R_{0102}=-R_{0132}-R_{0231}-R_{3132}
\end{cases},
\end{align}
occurs in terms of $R_{i0j0}$ that
\begin{align}
\Psi_{2}\left(\tilde{u}\right)=&-\frac{1}{6}R_{0303}+\mathcal{O}\left(\gamma_{k}\right),\label{352}\\
\Psi_{3}\left(\tilde{u}\right)=&-\frac{1}{2\sqrt{2}}R_{0301}+\frac{1}{2\sqrt{2}}iR_{0302}+\mathcal{O}\left(\gamma_{k}\right), \label{353}\\
\Psi_{4}\left(\tilde{u}\right)=&-\frac{1}{2}R_{0101}+\frac{1}{2}R_{0202}+iR_{0102}+\mathcal{O}\left(\gamma_{k}\right), \label{354}\\
\Phi_{22}\left(\tilde{u}\right)=&-\frac{1}{2}R_{0101}-\frac{1}{2}R_{0202}+\mathcal{O}\left(\gamma_{k}\right),\label{355}
\end{align}
while, in term of metric perturbation, the NP~scalars reads
\begin{align}
\Psi_{2}\left(\tilde{u}\right)=&-\frac{1}{12}\left(2h_{03,03}-h_{00,33}-h_{33,00}\right)+\mathcal{O}\left(\gamma_{k}\right),\label{356}\\
\Psi_{3}\left(\tilde{u}\right)=&-\frac{1}{4\sqrt{2}}\left(h_{01,03}-h_{13,00}\right)+\frac{1}{4\sqrt{2}}i\left(h_{02,03}-h_{23,00}\right)+\mathcal{O}\left(\gamma_{k}\right),\label{357}\\
\Psi_{4}\left(\tilde{u}\right)=&\frac{1}{4}\left(h_{11,00}-h_{22,00}\right)-2ih_{12,00}+\mathcal{O}\left(\gamma_{k}\right),\label{358}\\
\Phi_{22}\left(\tilde{u}\right)=&\frac{1}{4}\left(h_{11,00}+h_{22,00}\right)+\mathcal{O}\left(\gamma_{k}\right)\ .\label{359}
\end{align}
From Eqs.~\eqref{304}, for a general congruence of non-null geodesics of GWs, we obtain, for a massless mode $\omega_{1}$ and $n-1$ massive modes $\omega_{j}$,  at $\mathbf{k}$ fixed to lowest order in $\gamma_{k}$, the following expressions
\begin{equation}\label{359.1}
\begin{aligned}
\Psi_{2}\left(\tilde{u}\right)&=\mathcal{O}\left(\gamma_{k}\right) ,\\
\Psi_{3}\left(\tilde{u}\right)&=\mathcal{O}\left(\gamma_{k}\right)\ ,\\
\Psi_{4}\left(\tilde{u}\right)&=-\omega_{1}^{2}\biggl[\Bigl(\epsilon^{(+)}\left(\omega_{1}\right)e^{i\omega_{1}\left(t-z\right)}+c.c.\Bigr)\\&\qquad\qquad\qquad\qquad\qquad-i\left(\epsilon^{(\times)}\left(\omega_{1}\right)e^{i\omega_{1}\left(t-z\right)}+c.c.\right)\biggr]+\mathcal{O}\left(\gamma_{k}\right)\ ,\\
\Phi_{22}\left(\tilde{u}\right)&=\frac{k_{z}^{2}}{8}\sum_{j=2}^{n}\tilde{A}_{j}\left(k_{z}\right)e^{i k_{z}\left(t-z\right)}+c.c.+\mathcal{O}\left(\gamma_{k}\right)\ ,
\end{aligned}
\end{equation}
according to
\begin{equation}\label{360}
\omega_{j}=k_{z}+\mathcal{O}\left(\gamma_{k}\right)\ ,
\end{equation}
in units  $c=1$. 
Thereby, from Eqs.~\eqref{359.1} follows that, to zeroth order in $\gamma_{k}$, the $\omega_{1}$ mode has helicity 2 because only $\Psi_{4}\neq 0$ while $\omega_{j}$ modes have helicity 0 because $\Phi_{22}\neq 0$.  However the quasi-Lorentz invariant $E(2)$ class of GWS generated by non-local gravitational theory $R+\sum_{h=1}^{n}a_{h}R\Box^{-h}R$ is $N_{3}$, because for a generic non-null wave the NP~quantities are $\Psi_{2}=\Psi_{3}=0$ and $\Psi_{4}\neq 0\neq\Phi_{22}$ to zeroth order in $\gamma_{k}$.  This means that  the presence or absence of all modes is observer-independent. The components of driving-force matrix $S(t)$ can be expressed in terms of the six electrics components $R_{i0j0}$,  as
\begin{equation}\label{360.1}
S_{ij}(t)\equiv R_{i0j0}\ ,
\end{equation}
and the six amplitudes with direction $\hat{\mathbf{k}}=\mathbf{e}_{z}$ as
\begin{equation}\label{360.2}
\begin{aligned}
p_{1}\left(\mathbf{e}_{z},t\right)&\equiv\Psi_{2}\ ,\\
p_{2}\left(\mathbf{e}_{z},t\right)&\equiv\text{Re}\;\Psi_{3}\ ,\\
p_{3}\left(\mathbf{e}_{z},t\right)&\equiv\text{Im}\;\Psi_{3}\ ,\\
p_{4}\left(\mathbf{e}_{z},t\right)&\equiv\text{Re}\;\Psi_{4}\ ,\\
p_{5}\left(\mathbf{e}_{z},t\right)&\equiv\text{Im}\;\Psi_{4}\ ,\\
p_{6}\left(\mathbf{e}_{z},t\right)&\equiv\Phi_{22}\ .
\end{aligned}
\end{equation}
In matrix form,  $S(t)$ i terms of six  polarization matrices $W_{A}(\mathbf{e}_{z})$ defined by 
\begin{equation}.\label{362}
\begin{aligned}
W_{1}\left(\mathbf{e}_{z}\right)=&-6\begin{pmatrix} 
0 & 0 & 0 \\
0 & 0 & 0 \\
0 & 0 & 1
\end{pmatrix}\ , & W_{2}\left(\mathbf{e}_{z}\right)=&-2\sqrt{2}\begin{pmatrix} 
0 & 0 & 1 \\
0 & 0 & 0 \\
1 & 0 & 0
\end{pmatrix}\ ,\\
W_{3}\left(\mathbf{e}_{z}\right)=&2\sqrt{2}\begin{pmatrix} 
0 & 0 & 0 \\
0 & 0 & 1 \\
0 & 1 & 0
\end{pmatrix}\ , & W_{4}\left(\mathbf{e}_{z}\right)=&-\begin{pmatrix} 
1 & 0 & 0 \\
0 & -1 & 0 \\
0 & 0 & 0
\end{pmatrix}\ ,\\
W_{5}\left(\mathbf{e}_{z}\right)=&\begin{pmatrix} 
0 & 1 & 0 \\
1 & 0 & 0 \\
0 & 0 & 0
\end{pmatrix}\ , & W_{6}\left(\mathbf{e}_{z}\right)=&-\begin{pmatrix} 
1 & 0 & 0 \\
0 & 1 & 0 \\
0 & 0 & 0
\end{pmatrix}\ 
\end{aligned}
\end{equation}
yields 
\begin{equation}\label{361}
S\left(t\right)=\sum_{A}p_{A}\left(\mathbf{e}_{z},t\right)W_{A}\left(\mathbf{e}_{z}\right)\ ,
\end{equation}
where the index $A$ ranges over $\{1,2,3,4,5,6\}$~\cite{ELL,ELLWW}.

Accordingly,  there are six polarizations modes: the longitudinal mode $p_{1}^{\left(l\right)}$, the vector-$x$ mode $p_{2}^{\left(x\right)}$, the vector-$y$ mode $p_{3}^{\left(y\right)}$, the plus mode $p_{4}^{\left(+\right)}$, the cross mode $p_{5}^{\left(\times\right)}$, and the breathing mode $p_{6}^{\left(b\right)}$. Here $p_{A}\left(\mathbf{e}_{z},t\right)$ are the amplitudes of the wave at the detector in the frame origin~\cite{AMA,BCLN,WGRE, MN}. Then, the six polarization amplitudes $p_{A}\left(\mathbf{e}_{z}, t\right)$ in terms of the NP~scalars  for our non-local waves read
\begin{equation}\label{masspolarampl}
\begin{aligned}
p_{1}^{\left(l\right)}\left( \mathbf{e}_{z}, t\right)&=\mathcal{O}\left(\gamma_{k}\right)\ ,\\
p_{2}^{\left(x\right)}\left(\mathbf{e}_{z}, t\right)&=p_{3}^{\left(y\right)}\left(\mathbf{e}_{z}, t\right)=\mathcal{O}\left(\gamma_{k}\right)\ ,\\
p_{4}^{\left(+\right)}\left( \mathbf{e}_{z}, t\right)&=-\omega_{1}^{2}\tilde{\epsilon}^{\left(+\right)}\left(\omega_{1}\right)e^{i\omega_{1}t}+c.c.+\mathcal{O}\left(\gamma_{k}\right)\ ,\\
p_{5}^{\left(\times\right)}\left(\mathbf{e}_{z}, t\right)&=\omega_{1}^{2}\tilde{\epsilon}^{\left(\times\right)}\left(\omega_{1}\right)e^{i\omega_{1}t}+c.c.+\mathcal{O}\left(\gamma_{k}\right)\ ,\\
p_{6}^{\left(b\right)}\left(\mathbf{e}_{z}, t\right)&=\frac{k_{z}^{2}}{8}\sum_{j=2}^{n}\tilde{A}_{j}\left(k_{z}\right)e^{i k_{z}t}+c.c.+\mathcal{O}\left(\gamma_{k}\right)\ .
\end{aligned}
\end{equation}
To first order in $\gamma_{k}$, from Eqs.~\eqref{masspolarampl} only the amplitudes $p_{4}^{\left(+\right)}$, $p_{5}^{\left(\times\right)}$  and $p_{6}^{\left(b\right)}$ survive, while others are suppressed.  Precisely $p_{4}^{\left(+\right)}$ and $p_{5}^{\left(\times\right)}$ correspond to amplitudes of plus and cross modes of frequency $\omega_{1}$ predicted by general relativity while the amplitude $p_{6}^{\left(b\right)}$ is associated to the $j$ transverse breathing scalar modes of frequency $\omega_{j}$, always to lowest order.  In a nutshell, the gravitational radiation for non-local gravity \mbox{$R+\sum_{h=1}^{n}a_{h}R\Box^{-h}R$} exhibits three polarizations, i.e.  two tensor and, generally, one scalar, all transverse and $n+1$ modes.  Specifically,  is always present the standard radiation that is the associated with two modes, $(+)$ and $(\times)$, massless 2-helicity transverse and tensor, of frequency $\omega_{1}$, governed by two degrees of freedom (d.o.f.) $\tilde{\epsilon}^{(+)}(\omega_{1})$ and $\tilde{\epsilon}^{(\times)}(\omega_{1})$.  In addition under certain conditions are present  $n-1$ massive 0-helicity transverse scalar modes of frequency $\omega_{j}$,  namely the breathing mode, governed by $n-1$ d.o.f. $\tilde{A}_{j}(k_{z})$. It is worth observing that the same approach can be used also for higher order theories of gravity both in Riemannian, teleparallel and non-metric framework.  For details see~\cite{CCC1, CCC2}.
\section{Tables}\label{table}
A classification of solutions corresponding to null and non-null gravitational waves in higher order non-local gravity under some constraints, are summarized in following Tables~\ref{tab1} and \ref{tab2}.  In the Table~\ref{tab2.1} instead, we point out  Lagrangians in which the scalar radiation is prohibited while in Table~\ref{tab2.2} are summarized the count of massive scalar waves.  The polarizations and helicities of GWs with their related modes, taking into account the constraints, are showed in the following Tables~\ref{tab3} and \ref{tab4}.  Finally, the main results of this paper on gravitational waves in  higher order non-local gravity are summarized in the Table~\ref{tab5}, where we assumed $\phi_{1}^{(0)}=\lambda_{1}^{(0)}=0$.

\begin{table}[!ht]
\caption{Non-null wavelike solutions in $\Box^{-n}$-gravity.\label{tab1}}
\centering
\begin{tabular}{@{}lll|ll@{}} \hline
\text{Gravitational}  & $B=6a_{1}$ &\text{Solutions}& $B\neq 6a_{1}$ & \text{Solutions}  \\ 
 Lagrangian &$k^{2}\neq 0$ & &$k^{2}\neq 0$&\\ \hline
$R+a_{1}R\Box^{-1}R$ & $\infty^{1}$ solutions &$k^{2}\in\mathbb{R}_{+}$ & no solution & \\
$R+\sum_{h=1}^{2}a_{h}R\Box^{-h}R$  & no solution &&one solution & $k_{2}^{2}=\frac{6a_{2}}{6a_{1}-B}>0$ \\
$R+\sum_{h=1}^{3}a_{h}R\Box^{-h}R$ & one solution & $k_{2}^{2}=\frac{a_{3}}{a_{2}}>0$&two solutions &$k_{2,3}^{2}=\frac{\pm\sqrt{3a_{2}^{2}-12a_{1}a_{3}+2a_{3}B}-3a_{2}}{B-6a_{1}}>0$\\
$\vdots$ &$\vdots$ &$\vdots$ &$\vdots$ &$\vdots$ \\
$R+\sum_{h=1}^{n}a_{h}R\Box^{-h}R$& $n-2$ solutions &$k_{2}^{2},\cdots,k_{n-1}^{2}>0$& $n-1$ solutions &$k_{2}^{2},\cdots,k_{n}^{2}>0$\\ \hline
\end{tabular}
\end{table}

\begin{table}[!ht]
\caption{Null wavelike solutions in $\Box^{-n}$~-gravity. \label{tab2}}
\centering
{\begin{tabular}{@{}ccc@{}} \hline
\text{Gravitational Lagrangian}   &\text{Solutions} &\\ \hline
$R+\sum_{h=1}^{n}a_{h}R\Box^{-h}R$& $k_{1}^{2}=0$ for all $n$&\\ \hline
\end{tabular}}
\end{table}

\begin{table}[!ht]
\caption{Lagrangians where the scalar radiation is forbidden.\label{tab2.1}}
\centering
{\begin{tabular}{@{}lcc@{}} \hline
\text{Gravitational}  & \text{Constraint} &\text{GWs} \\ 
 Lagrangian &&\\ \hline
$R+a_{1}R\Box^{-1}R$ & $B\neq 6a_{1}$  & $(+)$ and $(\times)$ tensor modes\\
$R+a_{1}R\Box^{-1}R+a_{2}R\Box^{-2}R$  & $B=6a_{1}$&$(+)$ and $(\times)$ tensor modes \\ \hline
\end{tabular}}
\end{table}

\begin{table}[!ht]
\caption{Number of massive transverse scalar modes with helicity $0$ in higher order non-local gravity.\label{tab2.2}}
\centering
{\begin{tabular}{@{}ccc@{}} \hline
\text{Constraint}  & \text{Number of modes or d.o.f.} &\text{Order of $\Box^{-n}$} \\  \hline
$B\neq 6a_{1}$ & $n-1$  & $n\geq 1$\\
$B=6a_{1}$  & $n-2$&$n\geq 2$ \\ 
$B=6a_{1}$ & $\infty$ & $n=1$\\
\hline
\end{tabular}}
\end{table}

\landscape
\begin{table}[!ht]
\caption{Polarizations and modes of gravitational waves in higher order non-local gravity for $B\neq 6a_{1}$.\label{tab3}}
\centering
{\begin{tabular}{@{}ccccccccc@{}} \hline
\text{Gravitational }  & \text{Conditions} & \text{Frequency} & \text{Polarization} & \text{Type} & \text{d.o.f.}& \text{Helicity} & \text{Mass}\\
Lagrangian&& &modes&&&&\\
\hline
$R+\sum_{h=1}^{n}a_{h}R\Box^{-h}R$\hphantom{0} & \hphantom{0}$n\geq 1$ & \hphantom{0}$\omega_{1}$ & transverse & tensor & 2 & 2 & 0\\
&$k_{1}^{2}=0$&&$(+),(\times)$ &&&&\\\\
 \hphantom{0} & \hphantom{0}$n\geq 2$ & \hphantom{0}$\omega_{2},\ldots,\omega_{n}$ & transverse & scalar & $n-1$ & 0 & $M_{2},\ldots,M_{n}$\\
 &$k_{j}^{2}\neq 0$&&$(b)$ &&&&\\ \hline
\end{tabular}}
\end{table}

\begin{table}[!ht]
\caption{Polarizations and modes of gravitational waves in higher order non-local gravity for $B=6a_{1}$.\label{tab4}}
\centering
{\begin{tabular}{@{}ccccccccc@{}} \hline
\text{Gravitational}  & \text{Conditions} & \text{Frequency} & \text{Polarization} & \text{Type} & \text{d.o.f.} & \text{Helicity} & \text{Mass}\\
Lagrangian&&&modes&&&&\\
%&&&&& &\text{Petrov Class}&&\\ 
\hline
$R+\sum_{h=1}^{n}a_{h}R\Box^{-h}R$\hphantom{0} & \hphantom{0}$n\geq 1$ & \hphantom{0}$\omega_{1}$ & transverse & tensor & 2 & 2 & 0\\
&$k_{1}^{2}=0$&&$(+),(\times)$ &&&&\\\\
 degenerate state \hphantom{00} & \hphantom{0}$n=1$ & \hphantom{0}$\omega\in\mathbb{R}_{+}$ & transverse & scalar & $\infty$ & 0 & $M\in\mathbb{R}_{+}$\\
  &$k^{2}\neq 0$&&$(b)$ && &&\\ \\
 &$n=2$&no solutions &&&&\\
 &$k^{2}\neq 0$&&&&&\\\\
 \hphantom{00} & \hphantom{0}$n\geq 3$ & \hphantom{0}$\omega_{2},\ldots,\omega_{n-1}$ & transverse & scalar & $n-2$ & 0 & $M_{2},\ldots,M_{n-1}$\\
 &$k_{j}^{2}\neq 0$&&$(b)$ &&&&\\ \hline
\end{tabular}}
\end{table}

\begin{table}[!ht]
%\begin{center}
\caption{Polarizations and modes of gravitational waves in  in higher order non-local gravity.\label{tab5} }
%\end{center}
{\begin{tabular}{@{}ccll@{}} \hline
\text{Gravitational}  & \text{Constraint} & \text{Gravitational wave} & \text{Properties}\\
Lagrangian& %$G=c=1$
&{\centering solutions}& \\
%&&&&& &\text{Petrov Class}&&\\ 
\hline
$R+a_{1}R\Box^{-1}R$& $a_{1}\neq \frac{1}{6}$ &two  $(+)$ and $(\times)$ modes $k_{1}^{2}=0$ & tensor, transverse,massless, $2$-helicity\\
&&&\\
$n=1$&&&\\
& $a_{1}=\frac{1}{6}$&two  $(+)$ and $(\times)$ modes $k_{1}^{2}=0$&tensor, transverse, massless, $2$-helicity\\
  &  & $\infty$  breathing modes $k^{2}\neq 0$ & scalar, transverse, massive, $0$-helicity\\\\\hline
  $R+\sum_{h=1}^{2}a_{h}R\Box^{-h}R$&$a_{1}\neq \frac{1}{6}$&two  $(+)$ and $(\times)$ modes $k_{1}^{2}=0$& tensor, transverse, massless, $2$-helicity\\
 &&one breathing mode $k_{2}^{2}\neq 0$ & scalar, transverse, massive, $0$-helicity \\
  $n=2$&&&\\
& $a_{1}=\frac{1}{6}$& two  $(+)$ and $(\times)$ modes $k_{1}^{2}=0$ &tensor, transverse, massless, $2$-helicity\\\\\hline
$R+\sum_{h=1}^{n}a_{h}R\Box^{-h}R$&$a_{1}\neq\frac{1}{6}$& two  $(+)$ and $(\times)$ modes $k_{1}^{2}=0$&tensor, transverse, massless, $2$-helicity\\
&&$(n-1)$ breathing modes $k_{j}^{2}\neq 0$ & scalar, transverse, massive, $0$-helicity\\
$n\geq 3$&&&\\
 & $a_{1}=\frac{1}{6}$& two  $(+)$ and $(\times)$ modes $k_{1}^{2}=0$&tensor, transverse, massless, $2$-helicity\\ 
 &&$(n-2)$  breathing modes $k_{j}^{2}\neq 0$ & scalar, transverse, massive, $0$-helicity\\\\\hline
\end{tabular}}
\end{table}
\endlandscape

\section{Conclusions}\label{finalremarks}
The most interesting result of this paper is the presence of transverse scalar gravitational waves under certain conditions in some orders of gravity, that being massive propagate at a speed $v_{g}$ below $c$. More specifically, after localizing the non-local action and deriving the related field equations by a variational principle,  we linearized them weakly perturbing the tensor and scalar fields, because suppose to study the radiation at very high distances from the sources that had emitted it. Therefore,  the linearized equations in vacuum in a suitable gauge were resolved and we got non-null gravitational waves $k^{2}\neq 0$.  Indeed,  while the tensor radiation predicted by Einstein is always present in gravity of any order under any constraint,  further scalar radiation is prohibited in $\Box^{-1}$~gravity for $B\neq 6a_{1}$ and in $\Box^{-2}$~gravity for $B =6a_{1}$, see Table~\ref{tab2.1}.  If $n\geq 2$ and $B\neq 6a_{1}$ instead, the non-local theory exhibits $n-1$ wavelike scalar solutions $k^{2}\neq 0$,  where each $\Box^{-1}$ introduces a single massive scalar solution. While if $n\geq 2$ and $B=6a_{1}$, a single $k^{2}\neq 0$ scalar mode is suppressed and the non-local model shows $n-2$ wavelike scalar solutions. It is crucial to point out that under constraint $B=6a_{1}$, the $\Box^{-1}$~gravity shows a degenerate behavior in the meaning that any massive scalar waves $k^{2}\neq 0$ is solution of linearized filed equations in vacuum and under further assumption $\phi_{1}^{(0)}=\Lambda_{1}^{(0)}=0$ which implies $a_{1}=1/6$, the non-local $\Box^{-1}$~action restricted to two-dimensional spacetime is reduced to the Polyakov effective action. The polarizations of this radiation were analyzed both by means of geodesic deviation equation and Newmann-Penrose formalism. Thus, a transverse breathing polarization was found  for each $j$-th modes corresponding to each angular frequency $\omega_{j}$ to lowest order in $\gamma$,  a parameter that takes into account the difference in speed between the slightly massive wave and the massless one. The massive waves have helicity equal to zero and belong to $N_{3}$ class, according to $E(2)$~invariant classification of Petrov.

It would be very interesting in the future to study also the production of gravitational waves and not only their propagation in higher order non-local gravity, even in theories non-curvature based such as those torsion or non-metricity based.  Being able to measure the further massive scalar breathing polarization could be a signature of the viable theory of gravitation.
\section*{Acknowledgments}
MC acknowledges the Istituto Nazionale di Fisica Nucleare (INFN) Sez. di Napoli, Iniziative Specifiche QGSKY,  and the Istituto Nazionale di Alta Matematica (INdAM), gruppo GNFM, for the support.

\end{document}